%% file: main.tex
\documentclass[sigconf,screen]{acmart}
\renewcommand\footnotetextcopyrightpermission[1]{}

\usepackage{tikz}
\usepackage{amsmath}
\usepackage{amsthm}
\usepackage[sets,operators]{cryptocode}
\usepackage{float}
\usepackage{hyperref}

\usepackage{algpseudocode}
\usepackage{algorithm}

\usepackage{subfiles}
\usepackage{enumitem}
\usepackage{xspace}
\usepackage{booktabs, multirow}
\usepackage{svg}

\newcommand{\handlemath}[1]{\relax\ifmmode #1\else $#1$\fi}

\newcommand{\system}{\textit{TrustRate}}
\newcommand{\pollbot}{\gamma}

\newcommand{\totalvoters}{\handlemath{n}}
\newcommand{\numberofpolls}{\handlemath{M}}
\newcommand{\userparticipation}{\handlemath{c}}

\newcommand{\hijackthreshold}{\handlemath{\rho}}
\newcommand{\hijackedpolls}{\handlemath{\theta}}
\newcommand{\obtainedvotes}{\nu}
\newcommand{\id}{\mathsf{ID}}
\newcommand{\psid}{\mathsf{PSID}}
\newcommand{\malicioususers}{\mathcal{M}}
\newcommand{\allmalicioususers}{\mathcal{N}}
\newcommand{\subsetaudienceP}{\mathcal{S}}

\newcommand{\dynamicthreshold}{\handlemath{\mathcal{W}}}

\newtheorem{theorem}{Theorem}[section]
\newtheorem{lemma}{Lemma}[section]

\newtheorem{defn}{Definition}[section]

\newcommand{\eg}{\textit{e.g., }}
\newcommand{\ie}{\textit{i.e., }}

\newcommand{\URS}{\handlemath{\mathsf{URS}}}
\newcommand{\eddsa}{\textsf{EdDSA}}
\newcommand{\RSA}{\handlemath{\mathsf{RSA}}}

\newcommand{\blockene}{Blockene\xspace}
\newcommand{\citizens}{\textit{Citizens}\xspace}
\newcommand{\politicians}{\textit{Politicians}\xspace}
\newcommand{\citizen}{\textit{Citizen}\xspace}
\newcommand{\politician}{\textit{Politician}\xspace}
\newcommand{\txpool}{\handlemath{\mathsf{tx\_pool}}}
\newcommand{\txpools}{\handlemath{\mathsf{tx\_pools}}}

\newcommand{\MinimalState}{\textsf{MinimalState}}
\newcommand{\safes}{\handlemath{\mathcal{S}}}

\newcommand{\baselinereviewsystem}{\handlemath{\mathcal{B}}}

\newcommand{\pollP}{$P$}
\newcommand{\pollID}{\handlemath{\textsf{PID}}}
\newcommand{\pollIDP}{\handlemath{\textsf{PID}^{(P)}}}

\newcommand{\pollInfo}{\handlemath{\textsf{PollContent}}}

\newcommand{\voteValue}{\handlemath{\mathcal{V}}}

\newcommand{\voteReview}{\handlemath{\mathcal{V}_r}}
\newcommand{\voteComment}{\handlemath{\mathcal{V}_c}}

\newcommand{\numvotesreq}{\textsf{n\_req}}
\newcommand{\numvotesreqP}{\handlemath{\textsf{n\_req}^{(P)}}}

\newcommand{\numvotesseen}{\textsf{n\_seen}}
\newcommand{\numvotesseenP}{\handlemath{P^\textsf{n\_seen}}}

\newcommand{\audience}{\textsf{Audience}}
\newcommand{\audienceP}{\handlemath{\textsf{Audience}^{(P)}}}

\newcommand{\apathyfrac}{\mathsf{a}}

\newcommand{\categoryID}{\handlemath{\textsf{Topic}}}

\newcommand{\voters}{\textsf{Voters}\xspace}
\newcommand{\votersP}{\handlemath{\textsf{Voters}^{(P)}}}

\newcommand{\creator}{\textsf{Creator}}
\newcommand{\creatorP}{\handlemath{\textsf{Creator}^{(P)}}}

\newcommand{\VRFpoll}{\handlemath{\mathsf{VRF}}}

\newcommand{\votingWindow}{\handlemath{B_{vw}}}
\newcommand{\seedwaitperiod}{\handlemath{B_{wait}}}

\begin{document}
\date{November 2022}

\title{\system: A Decentralized Platform for Hijack-Resistant Anonymous Reviews}

\author{Rohit Dwivedula}
\authornotemark[2]
\affiliation{%
  \institution{The University of Texas at Austin}
  \city{Austin, TX}
  \country{USA}
}
\email{rohitdwivedula@utexas.edu}

\author{Sriram Sridhar}
\authornotemark[2]
\affiliation{%
  \institution{University of California, Berkeley}
  \city{Berkeley, CA}
  \country{USA}
}
\email{srirams@berkeley.edu}

\author{Sambhav Satija}
\authornotemark[2]
\affiliation{%
  \institution{University of Wisconsin, Madison}
  \city{Madison, WI}
  \country{USA}
}
\email{ssatija@wisc.edu}

\author{Muthian Sivathanu}
\affiliation{%
  \institution{Microsoft Research}
  \city{Bangalore}
  \country{India}
}
\email{muthian@microsoft.com}

\author{Nishanth Chandran}
\affiliation{%
  \institution{Microsoft Research}
  \city{Bengaluru}
  \country{India}
}
\email{nichandr@microsoft.com}

\author{Divya Gupta}
\affiliation{%
  \institution{Microsoft Research}
  \city{Bengaluru}
  \country{India}
}
\email{divya.gupta@microsoft.com}

\author{Satya Lokam}
\affiliation{%
  \institution{Microsoft Research}
  \city{Bengaluru}
  \country{India}
}
\email{satya.lokam@microsoft.com}

\subfile{sections/abstract}

\maketitle
\pagestyle{plain} %

\begingroup
\let\thefootnote\relax\footnotetext{$\dagger$ Work done while at Microsoft Research, India.}
\endgroup

\subfile{sections/1_introduction}

\subfile{sections/2_polling_system}

\subfile{sections/3_building_blocks}

\subfile{sections/4_design}
\subfile{sections/5_design_blockene_app}
\subfile{sections/6_results}

\subfile{sections/7_discussion}

\subfile{sections/8_related_work}

\bibliographystyle{plain}
\bibliography{citations}

\newpage
\appendix
\subfile{appendix/1a-blind-sigs}
\subfile{appendix/2b-urs_performance}
\subfile{appendix/3c-topics}

\subfile{appendix/4d-apathy}

\subfile{appendix/5e-trustrateoverblockene}
\subfile{appendix/6f-security_model_offloads}
\subfile{appendix/7g-proofs}
\end{document}

%% file: sections/abstract.tex
\begin{abstract}
Reviews and ratings by users form a central component in several widely used products today (\eg product reviews, ratings of online content, etc.), but today's platforms for managing such reviews are ad-hoc and vulnerable to various forms of tampering and hijack by fake reviews either by bots or motivated paid workers.  We define a new metric called `hijack-resistance' for such review platforms, and then present \system, an end-to-end decentralized, hijack-resistant platform for authentic, anonymous, tamper-proof reviews.  With a prototype implementation and evaluation at the scale of thousands of nodes, we demonstrate the efficacy and performance of our platform, towards a new paradigm for building products based on trusted reviews by end users without having to trust a single organization that manages the reviews.
\end{abstract}

%% file: sections/1_introduction.tex
\section{Introduction}
\label{sec:intro}

Reviews and ratings by end-users play a fundamental role today in the ranking and distribution of content (\eg `likes' on social networks or Youtube videos),  the ranking and consequent revenue of businesses on e-commerce sites (\eg product reviews on Amazon), and even the rating and remuneration of contractors/employees within aggregators (\eg Uber, Doordash).  Despite their central and increasingly pervasive role in widely used products and services, reviews today are managed in an ad-hoc manner, with no transparency or verifiability or accuracy guarantees, and are often completely managed under the control of one organization.   

Due to the lack of transparency and verifiability, reviews are vulnerable to tampering or bias either by the organization managing the platform (to favor a subset of content, opinion, businesses, or employees) or by external abuse of the platform such as spam bots or coteries with vested interests~\cite{social-media-manipulation} such as ``social-media-army'' (a small number of users who are paid to write fake or polarized reviews) propagating a particular viewpoint. As these biased reviews masquerade as genuine reviews, such bias or tampering of review outcomes (with or without the collusion of the organization managing the reviews) is a serious threat to the viability and utility of these services.   Worse, given the implicit trust that users and products place on these reviews, they pose a serious danger of manipulating the collective societal beliefs and opinions with far-reaching implications for society.

In this paper, we present \system, an end-to-end platform for gathering and recording representative, tamper-proof reviews that truly reflect the collective opinion of the target audience, and is resistant to `hijacks' or selective tampering by either a particular organization or a small subset of heavily interested users who try to skew the reviews.     

\subsection{Properties of \system}
\label{sec:intro_properties}
\system\ has three key properties that we believe are essential to the trustworthiness of a review platform:  it is decentralized, hijack-resistant, and anonymous. 

{\bf Decentralized:} To prevent selective bias by a central organization, \system\ is built on top of \blockene \cite{blockene_osdi}, a scalable and lightweight blockchain system where members can participate as first-class citizens with just a smartphone; while any blockchain system would be suitable for ensuring decentralization, we believe the low cost of participation makes \blockene\ a good fit for \system, as we ideally need the end users to collectively control the platform and ensure its integrity. 

{\bf Hijack-resistant:} To prevent hijack of reviews by spam bots, \system\ piggybacks on the Sybil-resistant properties of \blockene, which ties identity to either a unique smartphone (by relying on the TEE in Android/iPhone) or to a widespread external identity provider.   Any review in \system\ can only be done by a registered \blockene\ member.  To prevent the hijack of reviews by a small number of motivated/paid users, we employ a novel notion of {\em dynamic randomized access control (DRAC)} by allowing only a small (\eg 200) randomly chosen subset of users to review on a particular item.    Such a randomized access control has several desired effects: 1) It increases the cost of hijacking by several orders of magnitude; 2) It minimizes the voting burden on the honest users, thus making the system hijack-resistant against a broad class of attacks; and 3) It ensures that the outcome of the reviews/ratings is faithfully reflective of the broader audience. 

{\bf Anonymous and Unique voting:} An important requirement of any review or voting platform is to allow the user to express their free and honest opinion about the entity being reviewed (\eg a news article, video, product, or a person) without fear of being linked to their real identity.  \system\ ensures \textit{content anonymity}. This ensures that a user's review cannot be linked to their TrustRate identity. To provide this {\em content anonymity} among the users randomly authorized to review an entity, \system\ uses a modified form of a cryptographic primitive called {\em Unique Ring Signatures (URS)}~\cite{log_urs}. 

Additionally, \system\ can also be configured to guarantee a stronger property of \textit{membership anonymity}. {\em Membership anonymity} ensures the unlinkability of \system\ identities to the real identity of the user. In settings where the identities in \blockene\ are based on an external identity provider (\eg, an organization like Uber giving out unique identities to each driver), \system\ additionally ensures that the organization/identity provider cannot link \system\ identities to the real identity of the user. For this, \system\ uses another cryptographic primitive called {\em blind signatures}~\cite{blind_sigs_chaum1982}. In both settings, \system\ ensures that an individual can express at most one review for a given entity; i.e., reviews are {\em unique}. 

\subsection{\system\ architecture}
The randomized access control for each entity being reviewed is key to the hijack-resistance of \system. \system\ uses a novel concept of {\em opinion committees} to achieve this. For every new entity requiring reviews, a cryptographically random subset of users is chosen by \system\ to express their opinion/vote. Similar to the per-block random sub-committee in the block commit protocol of systems such as \blockene~\cite{blockene_osdi} and AlgoRand~\cite{algorand}, opinion committees in \system\ are defined for every unique poll (\eg~every unique video, news article, or survey posted to \system).  A user is permitted to review an entity only if the {\em VRF} (a cryptographically generated, verifiable token) for the \{user, entity\} combination falls within a dynamically computed threshold; this threshold is automatically chosen by the system based on the number of active users and user availability/willingness to participate in reviews.  As \system\ is built on top of \blockene, it is important to observe that opinion committees are distinct and unrelated to the {\em block committee} for committing a block in \blockene; while there can be only one active block committee at a given time (i.e. the committee for the current block), there can be thousands of opinion committees active at the same time, one for each active poll being voted on through \system.

Central to the efficiency of \system\ is the {\em co-design} of the application platform with the blockchain architecture.  In particular, \system\ defines the notion of a {\em transaction group} which is respected all the way down at the blockchain validation/commit protocol.   While traditional blockchains view each transaction as an independent entity, \system\ ensures processing all transactions within the group as a single batch, to be committed within the same block. This batching provides two key benefits.  First, by batching all votes for a particular poll as part of a single block, \system\ amortizes the high cost of verifying URS by integrating with the batch-verify feature of URS.  Second, the transaction grouping allows \system\ to be trivially shardable to scale transaction throughput, as the global state pertaining to each poll is completely independent of other polls, but all votes of a poll are mapped to the same block in the same shard.  
In addition to batch verification for efficiency, our URS scheme improves on  ~\cite{log_urs} by making non-black-box modifications including, incorporating a new zero knowledge proof to prevent double-voting with non-binary and string-valued votes/reviews. 

\subsection{Generality of \system}
\label{subsec:intro_usecases}
\noindent We demonstrate the generality of \system\ with two widely different case studies.  The first case study builds around the notion of {\em disaggregated aggregators}, where we explore the question of how a content aggregator (\eg~Google News, YouTube) can be redesigned with \system\ to put in trustworthy, hijack-resistant user reviews to build more transparency and credibility to the platform.  Here, the voters will be members of the general public who are registered with \system.  For this usecase, \system\ provides {\em content anonymity}: the review/vote of a user cannot be directly or indirectly linked to the \system\ identity of the user who posted the vote. However, the fact that a user cast {\em some vote} for that poll (\ie, was part of the opinion committee for a poll) may still be detected. We refer to this application of our system as the \textbf{permissionless} use-case, since anyone is allowed to sign up as a user.

The second case study explores an enterprise setting where an organization such as Uber (or a traditional employer with a large employee base) wants to conduct surveys/polls with the goal of driving policy changes based on the outcomes.  Today, employers engage third parties to conduct these polls, but users have no guarantees of anonymity or even the outcomes of the poll; with \system, a company can conduct and release the results of polls on sensitive topics in a transparent and ``zero-trust'' manner, where employees can trust the results of the poll even if they do not trust the employer. For this scenario, \system\ provides both content anonymity and membership anonymity, by using {\em blind signatures}~\cite{blind_sigs_chaum1982}. More broadly, this second scenario can be thought of as a \textbf{permissioned} use-case since applications such as these have a closed, private set of users.

We formalize \textit{hijack resistance}, a measure of how difficult it is to hijack a review system, and explain how \system\ improves over existing review platforms. We build a prototype of \system\ and evaluate it on a cluster of 2200 VMs on Azure spread across three geographic locations.

%% file: sections/2_polling_system.tex
\section{Polling System: Definition and Notation}
\label{sec:polling-system}

\subsection{Defining the review/polling system}
\label{sec:def-polling-system}

We use the terms polling system and review system interchangeably. Analogously, we use the terms vote and review interchangeably. A polling system is a collection of {\em users} and {\em polls}. \noindent A poll \pollP{} consists of following attributes: 

\begin{itemize}[leftmargin=0.2in, itemsep=1pt]
    \item \pollIDP: A unique identifier.
    \item \creatorP: The user who creates the poll.
    \item \votersP: Set of users that are eligible to vote on the poll. 
    \item \votingWindow: Voting window; the time window in which selected users cast their vote. 
     \item \numvotesreqP: Number of votes required for this poll. %
\end{itemize}
Note that $\votersP \subseteq \audienceP$. Our polling system defines criteria for selecting \votersP\ from \audienceP\ such that $|\votersP| = \numvotesreqP$. 

In subsequent sections, when the poll $P$ is clear from the context, we drop the superscript $P$ and instead write \pollID, \voters, etc.

\subsection{Properties of polling system}
\label{sec:def-pollng-system-properties}

We define four desirable properties of a polling system -- decentralized, hijack-resistance, anonymity, and fairness.

\subsubsection{Decentralized} Informally, we say that a review system is decentralized if no single entity has control over membership in the system, collection of reviews, or storage of data. Traditional aggregators, such as Google News/ YouTube, are entirely controlled by one entity, making them centralized. 
Ideally, a \textit{decentralized} review platform is completely governed under the shared control of users participating in it. 

\subsubsection{Hijack Resistance}
\label{sec:hijack_cost_definition}
Qualitatively, the set of reviews on an entity (\eg a video or a news article) is said to be `hijacked' if the cumulative set of reviews is tampered with or skewed either by the platform itself  or due to fake reviews by external parties (either bots or a motivated set of radicalized or paid persons).  A review platform is {\em hijack-resistant} if the cost of such an attack is very high.  To increase the hijacking cost it is necessary to  prevent bots on the system, otherwise, the hijacking cost will be negligible.  Even then, the review platforms today that require a real account and have bot-validation checks in the form of CAPTCHAs suffer from a fundamental problem: the asymmetric power that hijackers possess due to two key aspects: \begin{itemize}
    \item \noindent{\bf Review bandwidth:}   In contrast to honest people who have limited bandwidth (\eg\ review a couple of items per day), hijackers have nearly infinite bandwidth as they are highly motivated to write reviews.

    \item \noindent{\bf Coordination:}  Because the army of motivated hijackers can co-ordinate amongst themselves, they can pick very specific items (\eg high-stake news articles) to hijack.  Honest reviewers on the other hand, naturally fragment their review bandwidth across numerous items.
\end{itemize}

Because of such asymmetric power, it only takes a small number of hijackers to perform targeted hijacking of reviews.  As an example, let us say $\obtainedvotes_P$ is the number of votes obtained on poll $P$. If the goal of an attacker was to capture say 50\% of the responses to the poll, they would need to use only $\obtainedvotes_P/2$ accounts or real users (depending on the access control in the platform), as they can selectively pick which polls they participate in to skew their votes. Here, the {\em hijacking cost} is only $\obtainedvotes_P/2$.

If a review platform takes away this asymmetric power, this implies that an attacker with an agenda would need a very large number of hijackers to achieve the desired outcome, thus significantly raising the hijacking cost.  In \S\ref{sec:random_sampling}, we describe how \system\ increases this hijacking cost by several orders of magnitude with a novel notion of {\em Dynamic Randomized Access Control}.  Next, we describe hijacking cost and hijack resistance more formally and analyze it for a baseline system where users are allowed to review all entities.

\begin{defn}[Hijacking cost]
\label{defn:hijack_cost}
 Let $0 < \hijackthreshold, \hijackedpolls, \epsilon < 1$ be constants. 
For some poll $\mathcal{P}$, let $\obtainedvotes_\mathcal{P}$ denote the number of reviews collected by the system for that poll. We say that the poll $\mathcal{P}$ is {\em hijacked} if $> \hijackthreshold\cdot\obtainedvotes_\mathcal{P}$ of the reviewers for $\mathcal{P}$ are under the control of the attacker. The {\em hijacking cost} of a polling system, $\pollbot$, is defined as the minimum fraction of reviewers that must be controlled by the attacker to hijack $\hijackedpolls$ fraction of the total polls in the system, except for $\epsilon$ probability.
\end{defn} 

\noindent Let $\totalvoters$ be the total number of users. Consider a time period $T$ (\eg\ a month) such that the total number of polls in this time span is $\numberofpolls$. Let us define a new variable,  $\userparticipation$, denoting reviewer bandwidth. Reviewer bandwidth (\userparticipation) is defined as the number of reviews each non-hijacked member produces on average during some time period $T$. \\

\noindent{\bf Baseline review system ($\baselinereviewsystem$)}: Let us consider a baseline system where everyone on the system (\ie all reviewers) is eligible to vote once on every poll under review. As one extreme example, such a system is hijack resistant with the following parameters: $\pollbot = \hijackthreshold$ and $\hijackedpolls = 1$. What this means is that an adversary will have to control $\hijackthreshold$ fraction of all users before they are able to hijack \textit{all} the polls in the system. In this scenario, if we define a poll as \textit{hijacked} if 50\% of reviews on it are controlled by adversaries (\ie\ $\hijackthreshold = 0.5$), the hijack cost $\pollbot$, also becomes 50\% of all users (since $\pollbot = \hijackthreshold$). If the adversary controls $ < \pollbot$ fraction of users, no polls are hijacked at all. However, for this particular scenario, the bandwidth of honest users $\userparticipation$ has to be very high - in particular, $\userparticipation = \numberofpolls$, meaning that every user participates in every poll, which is an unacceptably high workload for honest reviewers. 

Another scenario in this system is that every honest user, on average, participates only in some small number, say, $\userparticipation$ reviews in time $T$ (which captures common review systems today). Informally, if all polls have an equal number of votes from honest users, then each poll has $\kappa \approx \frac{nc}{M}$ honest reviews \footnote{This assumption can be easily removed probabilistically}. Now, if $\hijackthreshold = 0.5$, an adversary would only need to control $\kappa$ users to hijack all polls. This system is hijack resistant with very poor parameters -- namely, $\pollbot \approx \frac{\kappa}{\totalvoters}$ (for $\hijackedpolls=1$, $\hijackthreshold = 0.5$) \footnote{More generally, $\pollbot\approx\frac{\hijackthreshold}{1-\hijackthreshold}\cdot\frac{\kappa}{\totalvoters}$.}.

We will show in \S\ref{sec:system_properties_hijacking_cost} how \system\ can obtain much higher $\pollbot$ while keeping the number of reviews by honest users $\approx \kappa$ per poll (\ie  respecting the reviewing bandwidth of honest users).

\subsubsection{Anonymity} Informally, \textit{content-anonymity} says that within a poll, the vote of an honest user is ``hidden''. Membership anonymity says that users' true identities cannot be tied to the identities used in the polling system. In the organizational setting, this anonymity is provided even against the organization and identity provider. Each user has a real world identity - $\id$ (\eg legal name); and a polling system ID ($\psid$) they use to access \system\ - a username or public key.

\begin{defn}\label{defn:contentanonymity}[Content-Anonymity of polls] For a poll $P$, let \votersP~be the set of voters (set of $\psid$s) and let a probabilistic polynomial-time attacker ($\mathcal{A}$) control a subset of voters $\malicioususers \subseteq \votersP$. Then, for every vote $V$ cast by an honest user, $\mathcal{A}$ succeeds in guessing the identity (\ie\ $\psid$) of the user associated with $V$, with a probability at most $\frac{1}{|\votersP|-|\malicioususers|} + \epsilon$, where $\epsilon$ is negligible.
\end{defn}

\begin{defn}\label{defn:membershipanonymity}[Membership Anonymity] 
Let $\id_1, \cdots \id_\totalvoters$ be the collection of true identities of users and let $\psid_1, \cdots, \psid_\totalvoters$ be the collection of identities in the polling system. 
Let $\allmalicioususers$ be the set of users in this universe under the control of some polynomial-time adversary ($\mathcal{A}$). 
Then, for any honest user $\id_i$ (\ie\ user not in $\allmalicioususers$), the adversary $\mathcal{A}$ can determine $\id_i$'s identity in the polling system with probability at most $\frac{1}{n-|\allmalicioususers|} + \epsilon$, where $\epsilon$ is negligible.  
Further, in the organizational setting, membership anonymity is also required to hold against the identity provider who assists in establishing the identities of users on the polling system.
\end{defn}

\subsubsection{Fairness}
Informally, fairness says that the eligible voters of a poll capture the voice of the audience well. That is, each group in the audience is suitably represented among the set of eligible voters. This condition is trivially satisfied in baseline systems that allow every user to vote on every poll. We will show in \S\ref{sec:system_properties_anonymity} that \system\ also satisfies fairness.
\begin{defn}\label{defn:fairness}[Fairness]
Let \audienceP be the set of users that a poll $\mathcal{P}$ is targeted at. Let $\subsetaudienceP$ be a subset of \audienceP. Let \votersP be the eligible voters for $\mathcal{P}$. Then with overwhelming probability, the fraction of voters in \votersP who belong to $\subsetaudienceP$ must be within  $\frac{|\subsetaudienceP|}{|\audienceP|}(1 \pm \delta)$ for some constant $\delta > 0$.
\end{defn}
\noindent Appendix~\ref{app:proof_fairness_anonymity} calculates the value of this $\delta$ and shows that \system\ satisfies this property.

%% file: sections/3_building_blocks.tex
\section{Building Blocks}

Our overall scheme uses three main building blocks that are combined to achieve the goals described in \S\ref{sec:intro}: the Blockene platform ~\cite{blockene_osdi}, unique ring signatures~\cite{log_urs}, and blind signatures~\cite{blind_sigs_chaum1982,blind_sigs_bellare2003one}. %

The first building block is a blockchain, which functions as a trustless, distributed backend to which data can be written and read. 
While any blockchain (\eg Ethereum \cite{ethereum}, Algorand \cite{algorand}, Avalanche \cite{avalanche}, etc.) would be suitable for verifiability and decentralization, we choose Blockene, specifically because of its' low cost participation. 

\subsection{Blockene}
\label{sec:blockene_basics}

\blockene~\cite{blockene_osdi} is a blockchain platform that enables resource constrained compute nodes (such as cell phones) to participate in blockchain consensus as first-class participants. \blockene\ uses a split-trust model in which there are resource-constrained nodes running on mobile (called \citizens) and powerful nodes running on servers (called \politicians). While a majority ($\geq 75\%$) of \citizens are assumed to be honest, only as low as $20\%$ of \politician nodes need to be honest for the security guarantees of \blockene\ to hold.

Unlike traditional blockchains, \citizens in \blockene\ are almost stateless - they are incapable of independently verifying transactions or fetching data stored on the chain. This is because \citizens do not locally possess a copy of the entire blockchain's transaction history - however, they do locally store some minimal amount of information on the state of the blockchain - eg. block hash, Merkle root hash, list of users on the chain. Let us call this the $\MinimalState{}$. Additionally, peer-to-peer communication between \citizens, which is required to reach consensus on the state of the blockchain after a set of updates, also does not happen directly; all messages are transmitted via \politicians.

\blockene, through a set of interactive protocols, allows \citizens to verifiably query \politicians to fetch required data or perform updates to data stored on the blockchain, while still maintaining strong soundness guarantees. There are three key protocols that enable this to happen: \begin{enumerate}
    \item \textsf{GetLedger}: This protocol allows a \citizen who possesses minimal state from $k$ blocks ago (\ie $\MinimalState{}_{i-k}$) to verifiably fetch $\MinimalState{}_i$ from \politicians.
    \item \textsf{GSRead}: A \citizen that possesses $\MinimalState{}_i$ locally can efficiently and verifiably query information stored on the blockchain from \politicians.
    \item \textsf{GSUpdate}: This protocol allows any \citizen who possesses $\MinimalState{}_{i}$ along with a list of updates to be made to the state (i.e. transactions) to verifiably compute $\MinimalState{}_{i+1}$.
\end{enumerate}

\subsubsection{Safe Sample}
\label{sec:blockene_safe_sample}
A key theme underpinning the three primitives listed above is the notion of a \textit{safe sample}. Instead of asking one \politician for any piece of information, \citizens instead ask multiple random \politicians ($\approx 25$) for the same piece of information. Using this, a \citizen that communicates with even a single non-malicious \politician (out of the safe sample) is guaranteed, through the protocol, to have a consistent and correct view of the blockchain's current state.   

\subsubsection{\politicians\ and Blacklisting}
\label{sec:blockene_politician_blacklisting}
There are a small number of \politicians\ ($\approx 200$), and each of these is run by either a corporation, a governmental organization, or a non-profit. As such, any \politician\ performing malicious behavior that can be succinctly \textit{provable} to third parties could immediately be \textit{blacklisted}. For example, a \politician\ sending differing block-hashes for a round when queried for it could easily be exposed by publishing both the signed messages with differing content, revealing the malicious behavior. \blockene's protocol uses such blacklisting mechanisms in multiple places to limit the kinds of attacks \politicians\ can perform. The severe reputational risks to the organization hosting a malicious politician, the constrained ability of nodes to be malicious due to blacklisting, and the fact that only a low fraction of \politicians\ ($\approx 20\%$) need to be honest for security guarantees to hold, makes it likelier that sufficient honest \politicians\ are available.

\subsubsection{Block Committees}
\label{sec:blockene_block_committees}
In \blockene, updates to the state of the blockchain are made by a randomly chosen committee of \citizens, called the \textbf{block committee}. Members of the committee first perform a \textsf{GSRead} to download the current state of the blockchain. \textsf{GSRead} is designed to ensure that committee members do not need to download the entire data on the chain - only the specific segments of data which will be updated in this block are selectively downloaded. Then, committee members locally compute the new state via the \textsf{GSUpdate} protocol. Once the \textsf{GSUpdate} protocol is complete, all committee members sign the new $\MinimalState{}_i$ that they just computed and publish it to all \politicians. Other \citizens, who were not part of the block committee use these signatures, as a part of the \textsf{GetLedger} protocol to verifiably download the new \MinimalState{}. 

While the authors of \blockene~\cite{blockene_osdi} implemented and reported results from their protocol being used as a generic transaction-based system, the protocol and security guarantees themselves are agnostic to what kind of data is being stored on the blockchain. Their implementation modeled the data being stored on the blockchain as a set of key-value pairs of integers. We implemented a version of the \blockene\ protocol that allows arbitrary data to be stored in the global state, allowing us to use \blockene\ as a layer in our system. 

Building on top of \blockene\ allows \system\ to be decentralized, tamper-proof, and verifiable. Additionally, \blockene's split trust design enables users of online services (who will typically be participating from smartphones) to control the blockchain without paying huge network, storage, or compute costs.

\subsection{Unique Ring Signatures}
\label{sec:building_blocks_urs}
Ring signatures~\cite{ring_sign} are digital signatures that allow a member of a group, called a ``ring'' to sign messages on behalf of the group without revealing which member of the ring is the actual signer. Prior works have also used these to provide anonymity in blockchain contexts~\cite{log_urs}. While there are many variants of ring signature schemes \cite{linkable_sign, fz_urs}, we use a \emph{Unique} Ring Signature (URS) scheme that builds on the one described in \cite{log_urs} with some non-trivial modifications (discussed in \S\ref{sec:urs_customisations}) to suit our application as well as to improve performance. The uniqueness property is crucial for our application to prevent double-voting (same user providing multiple reviews for the same entity). A second crucial property we need is compactness of signatures - this scheme satisfies this requirement as signature sizes growing logarithmically with the ring size.

\medskip

\noindent\textbf{Syntax.} A \URS{} scheme consists of four algorithms: $\mathsf{URS.Setup}$, $\mathsf{URS.KeyGen}$, $\mathsf{URS.Sign}$ and $\mathsf{URS.Verify}$. 
$\mathsf{URS.Setup}$ is a one time setup to create a set of public parameters ($URS_{pp}$) that are input to all other algorithms. $\mathsf{URS.KeyGen}$ for a user $u$ creates a secret key ($sk_u$) and public key ($pk_u$). The ring $R$ is defined as any set of public keys, i.e. $R = \{pk_1, pk_2, \ldots pk_{|R|}\}$. $\mathsf{URS.Sign}$ takes a message $M$, a ring $R$, and a secret key ($sk_u$) to generate a signature $S$. Finally, $\mathsf{URS.Verify}$ checks if the generated signature $S$ is valid for the ring $R$ and message $M$. 

In \URS{} schemes, a part of the signature is referred to as the {\em tag}, denoted by $\tau$, that is crucially used for "uniqueness" property. In \cite{log_urs}, the tag is generated using a pseudo-random function, denoted by $\mathsf{Tag}(\cdot)$, of message $M$, ring $R$ and secret key of the user $sk_u$. Specifically,
\begin{align}
    S = (\tau, \sigma), \text {where }
    \tau = \mathsf{Tag}(M, R, sk_u) = \mathsf{H}(M || R)^{sk_u}, 
\end{align}
\noindent where  $\mathsf{H}:\{0,1\}^* \rightarrow \mathbb{G}$ is a hash function. As is clear from the expression, there is a (cryptographically) unique tag for each tuple of message, ring and secret key. 

\noindent To summarize, our \URS{} scheme satisfies:
\begin{enumerate}[leftmargin=0.2in, itemsep=1pt]
    \item \textbf{Correctness:} For all messages $M$, rings $R$, an honestly generated signature by any user in $R$ is valid.
    \item \textbf{Non-Colliding: } For any message $M$, ring $R$, tags in the signatures of two distinct users from $R$ are distinct.   
\end{enumerate}

\newcommand{\adv}{\mathcal{A}}

\noindent\textbf{Security.} 
For security, we consider an adversary $\adv$ that controls a set of signing keys and gets to observe a collection of signatures by honest users under different rings. 

\begin{enumerate}[leftmargin=0.2in, itemsep=1pt]
    \item \textbf{Unforgeability:} If $\adv$ does not control any of the signers in a ring $R$, then it cannot produce a  signature on any message that is valid under $R$.
    \item \textbf{Anonymity:} It is infeasible for $\adv$ to guess (better than random) the real signer of a signature among the set of honest public keys in a ring.
     \item \textbf{Uniqueness:} For any messages and any ring, if $\adv$ controls $k$ signers (i.e., picks, possibly malicious $k$ signing keys) and produces $k+1$ valid signatures, then at least 2 of these signatures must have same tag (and hence, can be efficiently detected).
\end{enumerate}

The scheme in \cite{log_urs} satisfies the above properties in the random oracle model under the Decisional Diffie-Hellman (DDH) assumption~\cite{ddh_assumption}. Additionally, for this scheme, the size of the signatures only grows logarithmically in the size of the ring, enabling us to scale to large rings. Using \URS{} allows \system\ to de-link people's votes from their \URS{} public key, enabling anonymous participation and guarantees \emph{content anonymity} (cf. Definition~\ref{defn:contentanonymity}).

\subsection{Blind Signatures}
\label{sec:building_blocks_rsa_blind}

Blind signatures allows \system\ to guarantee \textit{membership anonymity}, allowing users to plausibly deny participation in polls itself. They \cite{blind_sigs_bellare2003one, blind_sigs_chaum1982} enable users to obtain signatures on a message of their choice without revealing the message itself to the signer. We use RSA-based blind signatures, which proceeds in a three step process: (a) the message to be signed is blinded by the user, (b) the signer signs this blinded message, and (c) the user leverages the properties of the blinding to reverse it and compute a signature on the original unblinded message. This process ensures that the signer cannot link the blinded message to the unblinded message and the signature. We defer formal definitions and properties of blind signatures to Appendix~\ref{app:blind-sigs}.

%% file: sections/4_design.tex
\section{\NoCaseChange{\system{} Design}}
\label{sec:design}
This section describes how the three main components of a polling system, namely, user registration (\S\ref{sec:user_registration}), poll creation (\S\ref{sec:create_poll}) and voting (\S\ref{sec:urs_customisations},   \S\ref{sec:random_sampling}) are designed to satisfy the properties described in \S\ref{sec:polling-system}. Then, we describe how we put all of the components together on top of \blockene\ to build the end-to-end polling system (\S\ref{sec:trustrate-over-blockene}).

\subsection{User Registration}
\label{sec:user_registration}
The permissioned and permissionless systems (\S\ref{subsec:intro_usecases}) use different registration mechanisms. In both scenarios, to register as a voter, a user needs to:
\begin{enumerate}[leftmargin=0.2in, itemsep=1pt]
    \item create a key-pair for unique ring signature ($\URS$) scheme (\S\ref{sec:urs_customisations}) to be used in voting later.
    \item publish the \textit{public key} of URS along with proof of membership (permissioned setting); or TEE-based authentication (similar to \blockene) in the permissionless setting using RegisterVoter transaction (\S\ref{sec:trustrate-over-blockene}). Proof of membership or TEE-based authentication protect against Sybil attacks.
\end{enumerate}
We discuss the two settings in detail below.

\subsubsection{Permissioned setting, e.g., organizational polling}
\label{sec:rsa_blinding}
For the permissioned setting, we require a user to register onto the platform with a public key that is authenticated by a central authority, e.g., admin of the organization. To protect against a Sybil attack by rogue users we require that a user first presents proof of membership in the organization. Furthermore, we require that this authentication step provides a strong privacy guarantee for the user. In particular, the organization cannot link the published public keys to the actual employees of the organization. In the context of the full scheme, this ensures that no one can determine which real-world identities are selected to vote on a specific poll. 

To achieve this, we use RSA Blind signatures (Section~\ref{sec:building_blocks_rsa_blind}) with the admin of the organization as the signer. Every employee generates a key pair ($pk, sk$) for voting locally, where $pk$ is the public key and $sk$ is the secret key. The employee and the organization interact in a session of the blind signature protocol with the message $pk$. The employee also provides proof of membership in the organization, e.g., logging in with their work email. At the end, the employee learns a signature $\sigma$ on the public key $pk$, which is used to register on \system.

\subsubsection{Permissionless setting}
\label{sec:tee_based_registration}
In the permissionless setting or public use case (e.g., disaggregated aggregators), we require everyone who intends to sign up for the polling to already be registered as a  \blockene citizen. As discussed in Section~\ref{sec:blockene_basics}, every \citizen has an \eddsa{} keypair. Users register for voting by creating a new key-pair for \URS{} scheme and publish a message containing the newly generated $\URS_m$ public key along with a signature under their \citizen \eddsa{} key. Linking voter registration to \blockene citizenship allows us to inherit all properties that  \citizens have (\ie prevention of Sybil attacks etc.).

\subsection{Creating a Poll}
\label{sec:create_poll}
Any registered user on our system can create a poll. 
Individual polling systems could set additional policies on poll creation - restricting permissions to create a poll to a certain set of users, or setting a rate limit on the number of polls a user can publish based on specific requirements that might vary from application to application. A new poll contains: %
\begin{itemize}[leftmargin=0.2in, itemsep=1pt]
    \item unique poll ID, \pollID{}.
    \item description of the question and type of responses accepted (binary, numeric, etc).
    \item (implicit or explicit) description of \audience{} of the poll.
    \item minimum number of votes required, \numvotesreq{} $<\abs{\audience{}}$.
\end{itemize}

\subsection{Voting on a Poll}
\label{sec:urs_customisations}
Our voting/review protocols are based on a Unique Ring Signature (URS) scheme. While (unique) ring signatures are used in the past in blockchains for anonymity (\eg see ~\cite{log_urs} and references therein), our use of them poses new challenges and we need to make several non-trivial improvements to apply them in our context. First, the existing schemes can prevent double-voting in case of binary votes only (details below). We need to allow for non-binary votes and also string-valued reviews and still prevent a user from casting/writing multiple votes/reviews for a given poll. We do this by adding a (non-interactive) zero-knowledge proof of \emph{equality of discrete logs} to the protocol in ~\cite{log_urs}. Secondly, our implementation improves on the performance reported in ~\cite{log_urs} through the use of more efficient \emph{elliptic curve} groups, for which we report benchmarks in Appendix~\ref{app:urs_m_details}. 
Thirdly, while compactness of this modified scheme follows from ~\cite{log_urs}, given our resource-constrained participants, verification time of our URS needs to be optimized. We do this by designing efficient \emph{batch} verification of several signatures (from the same poll) that replaces expensive group operations with scalar operations based on the mathematical structure of these signatures (explained in detail in Appendix~\ref{app:batch_verification_details}). 

Recall that during the user registration step, a user picks and commits to a public key of the $\URS{}$ scheme. When voting on a poll, the ring $R$ consists of the \audience{} of the poll (specified by the poll creation step). Consider a poll with ID \pollID\ and a user with public key $pk$, signing key $sk$, and vote $\voteValue$. The user creates a signature $S$ by signing the message $M = (\pollID || \voteValue)$ using $sk$ and ring $R$. Now, the user publishes $(M, S)$. The security of the ring signature scheme ensures that the signed voter's public key $pk$ cannot be linked to \voteValue.

However, this black-box use of \URS{}~\cite{log_urs} does not prevent double voting. Since the message contains the vote value itself (and not just the PollID) and the uniqueness of ring signatures only guarantees uniqueness per message, a user can cast two votes $\voteValue_1$ and $\voteValue_2$ without being detected. This attack is not an issue in the case of binary polls, i.e., when there are only two possible vote values (e.g., $0/1$, yes/no, disagree/agree); this is since double voting is equivalent to abstaining from voting and does not change the aggregate score/tally.

However, this is a legitimate attack for non-binary polls, i.e., polls where votes can take more than 2 values (e.g., $1,2,3,4,5$ used in movie ratings), and using the regular URS scheme does not suffice.

To overcome this attack of multiple votes by the same user, we make a few changes to the \URS{} protocol (from \S\ref{sec:building_blocks_urs}). In particular, in the original scheme, the tag $\tau$ is generated as a function of the message $M$ that does not suffice to prevent double voting. We change the scheme to have an additional tag $\nu$ that only depends on the poll ID \footnote{We cannot use only the tag that depends on \pollID\ since we need to sign the vote value as well.}: \\
\centerline{ $\nu \equiv \mathsf{Tag}(\pollID, R, sk_u)$
}
\\
In essence, we can ensure that each user votes only once by generating two separate signatures under \URS{} scheme, one each for $(\pollID || \voteValue)$ and $\pollID$ and use the tag in the second signature to check for multiple votes. 

This alone, however, doesn't suffice because one needs to link these 2 signatures in a tamper-proof manner (without breaking anonymity) to avoid tampering with honest votes. A na\"{\i}ve way to stitch two signatures would be to provide a separate non-interactive zero-knowledge proof. However, we propose an approach that does significantly better in terms of computational cost and signature size by making a simple non-black box change to the signing protocol.

At a high level, the two tags in our scheme are \linebreak $\tau = H(M||R)^{sk_i}$ and $\nu = H(\pollID || R)^{sk_i}$ for a hash function $H$. In the scheme, we generate the tag $\nu$ as part of \URS{} and generate $\tau$ separately, followed by a proof using an inexpensive sigma protocol ~\cite{sigma_protocol} to prove the equality of discrete logarithms for $\tau$ and $\nu$ with bases $H(M || R)$ and $H(\pollID || R)$  respectively. This avoids doubling of the cost in the na\"{\i}ve approach using two separate invocations of \URS{} to link the two signatures. 

A more formal description of our URS scheme, its security properties, its performance optimization, and microbenchmarks on our implementation are given in Appendix \ref{app:urs_m_details}.

\subsection{Dynamic Random Access Control (DRAC)}
\label{sec:random_sampling}

To reduce the cost of participation of honest users as well as increase the hijacking cost of the system (\S\ref{sec:def-pollng-system-properties}), instead of requiring everyone in the audience to vote on a poll, i.e., $\audienceP = \votersP$, we select a random subset of users from \audienceP to qualify as \votersP, i.e., $\voters \subset \audienceP$. In particular, in the common case, $|\votersP| \ll |\audienceP|$. Without loss of generality, the main body of this paper considers the case of a polling system where the \audience{} for every poll is every registered user on the platform (see \S\ref{sec:main_body_topics} for discussion). 

To pick a random sample of users from the audience we use a Verifiable Random Function (\VRFpoll{})~\cite{vrf_1999}. Our system ensures that a random, deterministic $\mathsf{seed}$ is generated every block in a manner as defined in \textit{Algorand}~\cite{algorand}. This $\mathsf{seed}$ is used as one of the inputs to the \VRFpoll{} function. The \VRFpoll{} for user $u$ on a poll \pollP{} with poll ID \pollID\ committed in block $i$ of the blockchain is defined by: 
\begin{equation}
    \label{eqn:vrf_ring_selection}
    \VRFpoll = H(\mathsf{seed}_{i+\seedwaitperiod}||\pollID||pk_u)
\end{equation}

\noindent where $pk_u$ is the public key of user $u$ and $\seedwaitperiod$ is a constant parameter\footnote{The parameter $\seedwaitperiod$ determines the number of blocks  users need to wait after a poll is committed before being able to  calculate the \VRFpoll{}s and determine the eligibility to vote. As soon as $\textsf{seed}_{i+\seedwaitperiod}$ is available, the set of users eligible to vote on the poll, \votersP, can be computed by anyone. See Lemma~\ref{thm:waiting_period} for details on how \seedwaitperiod\ is calculated to be $38$ blocks}. Based on this \VRFpoll{}, we select the \numvotesreq{} highest \VRFpoll{}s to be the reviewers for this poll, i.e., \votersP{}.

\subsubsection{Handling User Interests}
\label{sec:main_body_topics}

Instead of targeting all polls at all users, it is often desirable (and sometimes necessary) for a polling system to support targetting a specific poll at a certain subset of users. For example, in the public usecase, users could follow topics they were interested in to ensure that the polls they are randomly assigned are on issues that they enjoy perusing. In the permissioned setting, categories could be useful to group employees by their job roles. Our system as described so far is readily extensible to support multiple topics; when a new poll in a certain topic is created, instead of calculating the VRF (Eqn~\ref{eqn:vrf_ring_selection}) for all users, compute it only for the subset of users who are \textit{subscribed} to that topic and shortlist the top $\numvotesreq$ from that. Note that supporting multiple topics does cause some subtle differences in user registration and other aspects of the system - see Appendix \ref{app:handling_topics}.

\subsection{\system\ Properties}
\subsubsection{Hijacking Cost of \system}
\label{sec:system_properties_hijacking_cost}
Next, we discuss how our design significantly increases the hijacking cost of the polling system compared to the baseline systems used today. Recall  that the baseline system allows everyone to vote on all polls and the  honest users have limited bandwidth for participation, giving disproportionate power to malicious users who can participate in all polls. 
In contrast, our design of dynamic random access control restricts the participation of malicious users and makes the participation of malicious users similar to bandwidth of honest users. Hence, with this design, number of users that need to be corrupted to hijack the system is much higher. For the current discussion, we assume that every \textit{honest} user selected to participate in a poll, does so. This is a reasonable assumption since every user is selected to participate only a limited number of times. In Appendix~\ref{app:dynamic_thresholding}, we examine how the hijacking cost changes if a certain fraction of honest users exhibit \textit{apathy}, and decide not to participate even when selected to do so.

\begin{lemma}
\label{lemma:hijacking_cost_main_body}
Let $0< \hijackthreshold, \hijackedpolls, \epsilon < 1$ be constants as defined in \S\ref{sec:def-pollng-system-properties} and let $\gamma$ be used to denote hijack cost as defined in Defn~\ref{defn:hijack_cost}. Assuming that every honest user selected to take part in a poll does participate, \system\ is a review system with hijacking cost
$$\pollbot \coloneqq \frac{\hijackthreshold}{1 + g(\hijackthreshold \numvotesreq, \frac{\hijackedpolls}{1+g(\hijackedpolls \numberofpolls, \epsilon)})}$$ where $g(\alpha, \beta) = \frac{-3\ln \beta + \sqrt{(\ln \beta)^2 - 8\alpha \ln \beta}}{2\alpha + 2\ln \beta}$, $\numberofpolls$ is the number of entities reviewed in the system over some time period $T$ and $\numvotesreq$ is the number of unique reviews solicited for each poll.
\end{lemma}
\noindent Proof of Lemma~\ref{lemma:hijacking_cost_main_body} can be found in Appendix~\ref{lemma:hijacking_cost}.\\

\noindent For example, in a certain time window (\eg\ one month) that had $\numberofpolls = 10^6$ polls (with $\numvotesreq=50$ and $\totalvoters=10^6$), let us say we designate a poll as hijacked if 50\% of reviews in it are controlled by malicious users ($\hijackthreshold=0.5$). The cost to hijack 1\% of polls ($\hijackedpolls=0.01$) in our system is $\pollbot\approx0.237$, or 237,000+ people. Additionally, the reviewer bandwidth needed for our system is $\userparticipation\approx\frac{\numberofpolls\cdot\numvotesreq(1-\pollbot)}{\totalvoters}$, which for this configuration yields $\userparticipation\approx38.1$. A baseline review system, \baselinereviewsystem\ (\S\ref{sec:hijack_cost_definition}), with similar number of users, polls and reviewer bandwidth ($\userparticipation$) could be hijacked by only 39 motivated hijackers. This result is also shown in row 1 of Table~\ref{tab:hijack_cost_comp}, which repeats this calculation for other values of $\numvotesreq$, $\hijackthreshold$ and $\hijackedpolls$.

\begin{table}[!htp]\centering
\caption{A comparison of how many users an adversary needs to control to hijack each system, for a certain value of $\userparticipation$. We set $\numberofpolls=\totalvoters=10^6$ and $\epsilon=2^{-30}$.}

\label{tab:hijack_cost_comp}
\scriptsize
\begin{tabular}{lcc | c | cc}\toprule
\multicolumn{3}{c|}{\system\ Parameters} & &\multicolumn{2}{c}{\# of hijacked users needed} \\
\numvotesreqP &$\hijackthreshold$ &$\hijackedpolls$ & \userparticipation & \system\ & Baseline \\\midrule
50  &0.5 &0.01  & 38.1  & 237,521 & 39 \\
100 &0.5 &0.01  & 69.4  & 305,977 & 70 \\
200 &0.5 &0.01  & 128.3 & 358,413 & 129 \\ \\
50  &0.3 &0.01  & 44.7  & 105,398 & 20\\
100 &0.3 &0.01  & 84.5  & 154,328 & 37 \\
200 &0.3 &0.01  & 161.5 & 192,730 & 70 \\ %
\bottomrule
\end{tabular}
\end{table}

\subsubsection{Anonymity and Fairness of \system}
\label{sec:system_anonymity_and_fairness}
\noindent Since all participants of \system\ communicate via their mobile phones over the internet, malicious entities (either \politicians\ or other third parties) could potentially deanonymize voters by associating vote messages with the IPs they originated from. Additionally, once such a mapping of \system\ identities and IP addresses exists, malicious entities could also attempt to intimidate or bribe voters selected for a poll by sending them a message. However, we believe these concerns are not critical enough to compromise the anonymity of our system. Firstly, ``sending messages'', with just the IP address is not a straightforward task thanks to carrier-grade NAT~\cite{nat} and the way mobile operating systems design and restrict the use of push notifications. An untampered \system\ app installed on a mobile phone could also ignore any unrecognized traffic arriving on any existing channels. Additionally, other techniques to hide source IP such as onion routing~\cite{onion_routing} are also composable with our system to provide an extra layer of security. We prove the following properties in Appendix~\ref{app:proof_fairness_anonymity}. 

\label{sec:system_properties_anonymity}

\begin{theorem}
\label{thm:-anonymity-fairness}
The \system\ system satisfies fairness (Defn. \ref{defn:fairness}), content anonymity (Defn.~\ref{defn:contentanonymity}), and in the permissioned setting, membership anonymity (Defn~\ref{defn:membershipanonymity}). 
\end{theorem}

\subsection{Threat Model}
In our threat model, similar to \blockene, we assume that atleast $20\%$ of \politicians and $75\%$ of all \citizens are honest. The remaining $80\%$ of \politicians and $25\%$ of \citizens can be malicious and can collude with each other. 
One distinction to note is that malicious users in \blockene \citizens and malicious users in the definition of hijacking cost (Defn.~\ref{defn:hijack_cost}) do not necessarily refer to the same set of users or the same set of behaviours
\footnote{
Malicious behaviour in \citizens and \politicians primarily involves attacks on liveness or soundness of the block commit protocol. 
Malicious behaviour in the \system\ polling system involves skewing the votes on the polls themselves to paint an agenda or promote a specific cause. 
}. 
In \system's permissionless use case, every \citizen is uniquely linked to a $\URS_m$ key and both these sets of users are the same.  
However, in the permissioned setting, the users registering to vote on \system\ and the \citizens could possibly be differing sets of users.

%% file: sections/5_design_blockene_app.tex
\section{\NoCaseChange{Building \system\ over \textit{Blockene}}}
\label{sec:trustrate-over-blockene}
As discussed earlier, \system\ is built as an application over a blockchain - specifically, \blockene. 
In \system\, at a very high level, there are three major types of operations:
\begin{itemize}[leftmargin=0.1in, itemsep=0pt]
    \item $\textsf{RegisterVoter}$:  In this operation, user identities are registered on the blockchain through a transaction. While doing so, a well-formed URS public key, $\URS_{pk}$, is committed publicly. In the permissioned use-case, the block committee verifies that $\URS_{pk}$ has been certified by the organization's  signing key using the blind signature scheme of \S\ref{sec:building_blocks_rsa_blind}. In the permissionless setting use-case, the block committee verifies that $\URS_{pk}$ has been signed by a Blockene citizen and that there is only one $\URS_{pk}$ associated with each citizen. 
    \item $\textsf{CreatePoll}$ and $\textsf{CreateVote}$: Detailed information on the syntax of these operations is in Appendix~\ref{sec:txn_syntax}.
    For \textsf{CreateVote} requests, the block committee first checks if the poll is still open and accepting responses (\ie still within the voting window). 
    Then, they compute the list of users selected to participate in that poll (i.e. compute the ring $\mathcal{R}$), and verify if the signature $S$ is valid on $\URS_m$ and $\mathcal{R}$. All these checks are compute-intensive, and we discuss how we optimize this verification process in \S\ref{sec:opt}.
\end{itemize}
Additionally, to \textit{read} existing votes from our system, users use the \textsf{GSRead} protocol (\S\ref{sec:blockene_basics}) of \blockene since all poll and vote related data is committed to the global state. 

\subsection{Design Optimizations}
\label{sec:opt}
In \system, identifying the list of eligible voters via the \VRFpoll\ function (\S\ref{sec:random_sampling}) and verifying $\URS_m$ votes (\S\ref{sec:urs_customisations}) are both computationally expensive tasks. Building \system\ on an unmodified version of \blockene\ would require both these steps to be performed by \citizens, since \citizens are responsible for verifying signatures, checking logical correctness, and computing the updated global states. 

Building on key ideas from \blockene, such as that of the safe samples (\S\ref{sec:blockene_safe_sample}) and blacklisting (\S\ref{sec:blockene_politician_blacklisting}), we design protocols to significantly reduce the time required to complete these two expensive steps. We implement these optimisations, which represent non-black box use of the \blockene protocol, and quantitatively discuss the impact of these optimisations in \S\ref{sec:design-optimisation-impact}. Formal discussion on the threat models associated with these changes can be found in Appendix~\ref{app:optimisations_security_analysis}.

\subsubsection{Computing the Ring/Eligible Voters}
\label{sec:opt_computing_ring}
A key step for a block committee is to determine the set of eligible voters (\ie the ring) for new polls which are open for voting. Once this is calculated, the block committee adds the hash of the ring (\textit{ring-hash}) to the global state, to allow future users (and block committees) to retrieve the ring when needed.

For a poll \pollP, a committee member needs to do $N$ \VRFpoll\ hash computations where $N = |\audienceP|$. 
In the public use case, $N$ can easily be of the order of 1 million. Moreover, each block is expected to have around $\approx40$ polls.\footnote{With ring size $100$, each $9$~MB \blockene\ block has $\approx 4000$ transactions, out of which we expect $1\% (\approx 40$) to be new polls in the steady state.}

Our microbenchmarks indicate that performing $10^6$ hashes takes around $4 - 8$s on Android mobiles, depending on the make and model of the phone. Performing $40M$ hashes, for all $40$ polls in a block, would take around $\approx 160$ seconds, which is almost 2x the block commit time of 89s reported in \blockene\ \cite{blockene_osdi}. To alleviate this, we offload the task of computing the ring to the \politicians in a verifiable manner. 

All \citizens in the committee begin by querying a safe sample $\safes$ of \politicians (used for other \blockene tasks as well) for the ring-hash. If all \politicians in $\safes$ respond with the same hash value, the \citizen accepts this hash to be correct, and proceeds to the next stage. If they receive differing hashes, they (1) ask all the \politicians in $\safes$ to send the list of UUIDs of users that make up the ring, (2) combine all the lists (union) received and calculate $\VRFpoll$s for all UUIDs, and (4) shortlist the top-$k$ $\VRFpoll$s to obtain the ring. This reduces number of hash computations from $N$ to $\safes \cdot k$, even in the worst case. (additional details: Appendix~\ref{app:security_ring_offload}).

\subsubsection{Verifying ring signatures}
\label{sec:opt_verifying_urs}
A block committee needs to verify the signatures on all the transactions being considered in a block. In the original \blockene\ scheme, the transactions were signed using EdDSA signatures which take $\approx 0.2$ ms per verification on a mobile phone. 
In our polling system, the bulk of the transactions are vote transactions, which are signed using ring signatures. 
Verifying a single $\URS_m$ signature for a ring size of 128, for example, takes 26ms (Table~\ref{tab:urs_sign_verify} in Appendix~\ref{app:batching_urs}); ${\approx}130\times$ higher than EdDSA signatures. Even if we assume perfect batching (Appendix~\ref{app:batch_verification_details}) were somehow possible (i.e. verifying all the votes of a poll together), the amortized cost of verification per signature is around 8 ms, or still $30\times$ higher than EdDSA signatures.  To remove this performance bottleneck, we off-load signature validation to \politicians, which verify signatures in the background.

A \citizen queries a safe sample of \politicians to check whether the signatures on all the transactions being considered are valid. If not, the \politicians\ return a list of invalid transactions that are then explicitly verified by the \citizen. In the common case, when a bulk of the transactions are valid, most of the verification gets off-loaded. If a \politician tries to slow down a \citizen by claiming correct transactions to be incorrect the \politician can be blacklisted.\footnote{All messages exchanged between nodes are authenticated by the sender. Hence, an incorrect claim by a \politician can easily be used to blacklist them.}

Note that \citizens do not even have to obtain the ring to verify most \URS{} vote transactions -- they can directly ask a safe sample of \politicians\ whether the signatures in a set of transactions are all correct. If any transaction is reported as incorrect, \citizens\ can ask a \politician\ for the ring of eligible voters for that poll – that \politician\ then provides a list of voters, along with proof.\footnote{Recall that for every new poll, the block committee commits the hash of the ring, $H(R)$, to the global state. Thus, any \citizen\ can query the global state (with Merkle Tree challenge path) to verify if the ring is correct.} (additional details: Appendix~\ref{app:security_vote_verification}).

\subsubsection{Batch verification of signatures}
\label{sec:opt_batchverify}

Offloading verification of votes, as described in the previous section, works by removing the vote verification from the critical path of execution, and delegating it to be done in the background on \politicians. The simplest way a \politician\ could implement this is by verifying each vote individually as soon as it receives it. However, batching multiple votes from the same poll, and using the batching protocol we describe in Appendix~\ref{app:batch_verification_details} allows us to reduce the cost of verification per signature significantly.

In \blockene, the decision on which transactions to include in the current block is made by a deterministic, random subset of $45$ \politicians every block. Each of these selected \politicians\ commit to a \txpool, a set of transactions UUIDs that they intend to include in the current block. To ensure that multiple \politicians don't commit to the same set of transactions, every transaction is also uniquely, deterministically mapped\footnote{In round $i$, allow only the $j^{th}$ \politician out of these $45$ to include a certain vote $\mathcal{V}$ in it's \txpool, where $j\equiv\mathsf{Hash}(i || \mathcal{V})\bmod45$} to one of these $45$ politicians, essentially sharding the pending transactions into disjoint sets. In such a system, even if a \politician had multiple votes from the same poll in its pool of waiting transactions, it would not be able to group and commit to all of them, since some votes might be in a different shard.

For \system, we change \blockene's transaction selection policy to respect the notion of \textit{transaction groups}. Votes for the same poll \pollID\ fall in the same \textit{transaction group}. Concretely, for round $i$, a politician $j$ can include a vote V iff $j = \mathsf{Hash}(i||\pollID) \bmod 45$. This change allows \politicians to effectively benefit from the performance gains of $\URS_m$ batch verification. Results from \S\ref{sec:results_batching_impact} demonstrate that verifying a batch, even as small as two signatures, is 30-80\% faster than serially verifying them. The gains become more significant at larger batch sizes and larger ring sizes. Honest \politicians can deploy various heuristics to determine when to commit to a transaction group. Waiting for votes to stream can significantly improve throughput at the cost of a few blocks of additional latency. (additional details: Appendix~\ref{app:security_batched_vote_verify}).

\subsection{Identifying eligibility in polls}
\label{sec:opt_eligibility_in_polls}
Periodically, users need to determine if there are any new polls they are eligible to vote for. Let us say a certain poll $\pollID$'s voting window begins in block $B_i$. The block committee of $B_i$ obtains the hash of all members in the ring for that poll (Equation~\ref{eqn:vrf_ring_selection}) using the optimizations discussed and commits it to the GlobalState. \citizens\ obtain the list of new polls they are eligible for as follows:
\begin{enumerate}[leftmargin=0.1in, itemsep=0pt]
    \item Query \textit{all} \politicians\ asking them if they are any new polls in $B_i$ for which they are eligible. A malicious \politician\ could \textbf{drop polls}, saying there are no new polls the user is eligible for, even when that is false. However, this is easily detectable and blacklistable. 
    \item In case all \politicians\ return the same list of polls, no further action is needed, and we have the list of all polls we are eligible for. If there is a conflict in the lists provided, the \citizen\ calculates the union of all the \pollID{}s provided by the various SNs. For each of these \pollID{}s, the \citizen\ requests a randomly selected \politician\ to provide the list of users (ring) selected for that poll, the hash of the ring (which would have been committed to the GlobalState), and a Merkle challenge path to that hash. With this information, \citizens can locally determine which \politicians was being untruthful and blacklist it. 
    \item For each $\pollID$, download the ring and the poll and send the user a notification that there's a new poll available to vote on. (additional details: Appendix~\ref{app:security_determining_eligible_polls}).
\end{enumerate}

%% file: sections/6_results.tex
\section{Implementation Details \& Results}
\label{sec:implementation_and_results_section}

\subsection{Implementation}
We build our application atop \blockene's \cite{blockene_osdi} codebase, which consists of two major parts -- a C++ app (for the \politician) and an Android app for \citizens. The unique ring signature system is implemented as a C++ library and has $\sim1700$ lines of code with NTL~\cite{NTL} used for polynomial arithmetic operations and Sodium~\cite{libsodium} for finite field arithmetic. We use Android NDK~\cite{android_ndk} to compile this library for mobile devices, which can then be called from within the \textit{Citizen} Android app via JNI~\cite{jni}.

\subsection{Experimental Setup}
We simulate a \system\ instance with one million users who participate by sending in votes when selected. By default, we use a ring size of $100$, \ie hundred \system\ users are selected to vote in every poll\footnote{Performance numbers for other ring sizes are in Section~\ref{sec:results_ring_block_ablation}}. We continuously create and bombard the system with enough new polls (each of which is $\approx0.35$ KB) to ensure that \politicians' pending transaction pools are always sufficiently full. Each review consists of a 256-byte field - sufficient for both a numeric rating and a short text response, along with a signature of the reviewer. The signature size depends on the size of the ring being used (\eg with ring size 100, signature size is $\approx$ 1.95KB and the total size of content + signature is 2.2KB).

We use a setup similar to the one in \cite{blockene_osdi}. As discussed in \S\ref{sec:blockene_block_committees}, even if there are millions of registered \citizens, only $2000$ of them are actively participating in the \textit{block committee} at any point of time. As a result of this, even a \system\ instance with a million users can be simulated on a testbed of consisting of 2000 \citizen{} nodes and 200 \politician{} nodes, which is the experimental setup we used to perform experiments on Azure. 

Each \textit{Citizen} node is a 1-core, 2GB RAM virtual machine running an Android 7.1 image, with network rate-limited to 1MB/s. Each \textit{Politician} node is an 8-core VM (Intel Xeon Platinum 8370C, 8171M or 8272CL), with 32GB of RAM, running a Ubuntu LTS image, with network bandwidth rate limited to 40MB/s. \politicians\ are spread across two Azure regions -- EastUS and WestUS3, with 100 per region. \citizens\ are spread across three Azure regions -- SouthCentralUS (700), WestUS3 (600), and EastUS (700). For all experiments, we use the default \blockene{} block size of 9MB.

\subsection{Results}
\begin{itemize}[leftmargin=0.1in, itemsep=0pt]
    \item \S\ref{sec:results_perf_evaluation} reports system performance under increasing proportions of malicious nodes 
    \item \S\ref{sec:design-optimisation-impact} and \S\ref{sec:results_batching_impact} present the quantitative impact of our design optimisations (\S\ref{sec:opt})
    \item \S\ref{sec:results_anonymity_cost} examines the cost of \textit{content}/\textit{membership anonymity}
    \item \S\ref{sec:results_ring_block_ablation} presents throughputs for varying ring sizes.
    \item \S\ref{sec:results_participation_costs} analyzes participation cost for users in \system.
\end{itemize}

\subsubsection{Performance evaluation}
\label{sec:results_perf_evaluation}
We mimic malicious behavior on the blockchain similar to the evaluations of \cite{blockene_osdi}. Malicious nodes in the \blockene\ cause a reduction of throughput by dropping transactions, proposing and committing empty blocks, and wasting the network bandwidth of honest nodes by forcing them to resend more data during gossip. Additionally, in \system\, the optimizations we introduced (\S\ref{sec:opt}) provide additional attack surfaces for dishonest \politicians (discussed in Appendix~\ref{app:optimisations_security_analysis}) that allow them to cause stalls during the block committee. We report results for varying levels of malicious users in the \textit{Blockene} backend - ranging up to $25\%$ in \textit{Citizens} and $80\%$ in \textit{Politicians} (Table~\ref{tab:pollingapp_100_1_malicious_simulation}). In the fully honest (0\%/0\%) case, \system\ committed 132,930 votes and 1,342 articles in 35 blocks (2518 seconds), representing a total of $\approx 288$MB of data (votes + their signatures). Additionally, each block's latency is $71.94 \pm 4.84$ seconds. Extrapolating, \system\ can commit approximately $4.5$ million votes and $45,000$ polls per day in the steady state.

\begin{table}[htbp]\centering
\caption{Throughput of \system\ in \textbf{votes/sec} under varying levels of \citizen\ and \politician\ dishonesty. (ring size = 100)}
\label{tab:pollingapp_100_1_malicious_simulation}
\scriptsize
\begin{tabular}{lccc}\toprule
\multirow{2}{1cm}{Citizen Dishonesty} &\multicolumn{3}{c}{Politician Dishonesty} \\\cmidrule{2-4}
&0\% & 50\% & 80\% \\\midrule
0\%  & 52.79   & 28.55   & 14.54   \\
10\% & 39.20    & 24.69   & 13.04  \\

25\% & 31.43    & 17.15   & 11.17  \\
\bottomrule
\end{tabular}
\end{table}

\noindent Since votes in \system\ and transactions \blockene\ are significantly different in terms of size, the votes/sec of our system is not directly comparable to the 1045 txns/second throughput reported by the authors of \blockene. A value that is somewhat comparable is the total \textbf{data throughput} of both these systems, which is $\approx117$KB/s in our fully honest case and 114 KB/s in \blockene.

\subsubsection{Impact of offload optimizations}
\label{sec:design-optimisation-impact}
Here we examine how the throughput of \system\ increases with our optimizations of verifiably offloading computation and URS verification from \citizens to \politicians. In Table~\ref{tab:impact_of_optimisations}, configuration E1 represents an unmodified \blockene\ backend; \citizen nodes are completely responsible for verifying signatures and computing the new global state locally. In E2, \citizens\ offload the ring computation, a step required to calculate the new global state to \politicians and in E3, \citizens\ offload signature verification as well. From the table, it is evident that our customisations to the \blockene\ protocol (\S\ref{sec:opt}) improve the throughput by almost $4\times$. 

\begin{table}[htp]\centering
\caption{Impact of optimizations on vote throughput (ring size = 100)}
\label{tab:impact_of_optimisations}
\scriptsize
\begin{tabular}{ll}\toprule
Configuration &Throughput (votes/s) \\\midrule
E1: Baseline \blockene backend                            & 13.58    \\
E2: Offloading only ring calculation                      & 29.76    \\
E3: Offloading both ring calculation and URS verification & 52.79   \\
\bottomrule
\end{tabular}
\end{table}

\subsubsection{Impact of batching}
\label{sec:results_batching_impact}

We perform microbenchmarks on \politicians to demonstrate the impact batch verification of $\URS_m$ signatures can have\footnote{Note that all other results present E2E throughputs for the most pessimistic case where every vote needs to be individually verified with no batching.}.
Table~\ref{tab:microbenchmark_batching_impact_sn} presents results on time taken to batch-verify votes in rings ($\mathcal{R}$) of various sizes. 
From the table, we see that verifying even two signatures at once provides gains of $1.3$x at $\mathcal{R} = 50$ to $1.8$x at $\mathcal{R} = 1000$. 
For $\mathcal{R}=100$, each block has ${\approx}4000$ votes in it, and would take ${\approx}66$s to verify serially; batching would reduce this to $31.6$s with batch size 2 or $23.6$s with batch size 8. 

\begin{table}[htp]\centering
\caption{Time (ms) \textbf{per verification} when batch-verifying multiple votes}\label{tab:microbenchmark_batching_impact_sn}
\scriptsize
\begin{tabular}{lccccccccc}\toprule
\multirow{2}{0.5cm}{Ring Size} &\multirow{2}{0.5cm}{Single Verify} &\multicolumn{7}{c}{n\_votes\_batched} \\\cmidrule{3-9}
& &2 &3 &4 &5 &6 &7 &8 \\\midrule
50 &10.4 &7.9 &7.0 &6.5 &6.3 &6.1 &6.0 &5.9 \\
100 &16.5 &11.3 &9.6 &8.7 &8.1 &7.8 &7.5 &7.3 \\
200 &27.9 &17.5 &14.0 &12.2 &11.2 &10.5 &10.0 &9.6 \\
500 &60.8 &34.7 &25.9 &21.6 &18.9 &17.2 &16.0 &15.0 \\
1000 &115.2 &63.1 &45.5 &36.8 &31.6 &28.1 &25.6 &23.7 \\
\bottomrule
\end{tabular}
\end{table}

\subsubsection{Cost of anonymity} 
\label{sec:results_anonymity_cost}

In the previous section, we examined how our modifications to the \blockene protocol (and our design of $\URS_m$'s batch-verify) has improved the performance of \system\ by a significant factor. In this section, we attempt to quantify the additional \textbf{cost} of adding $\URS_m$ and blind signatures. Row M1 of Table~\ref{tab:result_cost_of_each_anonymity} is a baseline scenario\footnote{Note that the 262.75 votes/sec indicated for scenarios M1 and M2 \textit{includes} usage of the ring offload optimisation we developed in \S\ref{sec:opt_computing_ring}. Without this optimization, using a non-modified \blockene\ backend would result in a throughput of 23.8 votes/sec instead - more than $10\times$ lesser.} where there is no anonymity at all - users register to the system using a pseudonymous EdDSA key, and use that key for all interactions with the underlying \blockene protocol, as well as for casting votes on polls they are selected for. This key is linked either to their TEE (in the public use case) or to their employee ID (in the organisational case). M2 is a similar scenario, which can only be used for the permissioned organisational use-case, where blind signatures (\S\ref{sec:rsa_blinding}) are used to ensure EdDSA keys are not linkable to employee IDs.  

\begin{table}[htp]\centering
\caption{Cost of content and membership anonymity (ring size=100)}\label{tab:result_cost_of_each_anonymity}
\scriptsize
\begin{tabular}{lcc}\toprule
Configuration &votes/sec & Voting Scheme \\\midrule
M1: No anonymity at all                      & 262.75 & EdDSA \\ %
M2: Only membership anonymity (blind sigs)   & 262.75 & EdDSA \\ %
M3: Only content anonymity    ($\URS_m$)     & 52.79  & $\URS_m$ \\ %
M4: Content + membership anonymity           & 52.79  & $\URS_m$ \\ %
\bottomrule
\end{tabular}
\end{table}

We notice from Table~\ref{tab:result_cost_of_each_anonymity} that the system performance in (M1, M2) and (M3, M4) are the same -- adding membership anonymity requires a change \textit{only} in the setup phase. Once the blind signature protocol has been executed, the steady state throughput of the system depends on the voting scheme alone. Appendix~\ref{app:rsa_blinding_microbenchmarks} contains microbenchmarks that quantify the cost of this one-time setup phase. For some organisational settings, using a \system-like system with only membership anonymity might be a suitable option, since it is $5\times$ faster\footnote{The only downside in not having \textit{content anonymity} in a scenario like this is that adversaries will be able to build voting profiles of users, by compiling a list of polls each key voted on, and the corresponding value of that vote. Using $\URS_m$ would prevent attacks such as these.}.

\subsubsection{Varying ring sizes}
\label{sec:results_ring_block_ablation}

Most of the results so far have used a ring size of 100 for polls (\ie 100 users selected to vote on each poll). Table~\ref{tab:result_varying_ring_size} presents results on vote throughput for increasing sizes of rings. As the size of the ring increases, the signature size grows logarithmically, which in turn forces each block to hold a smaller number of votes.

\begin{table}[htp]\centering
\caption{\system\ throughput for various ring sizes. The maximum number of votes that can fit in each block varies since each block in \blockene\ is capped at a fixed size. }
\label{tab:result_varying_ring_size}
\scriptsize
\begin{tabular}{lccccc}\toprule
\textbf{Ring Size} &\textbf{Vote Size (bytes)} &\textbf{max\_votes} &\textbf{votes/sec} &\textbf{KB/sec} \\\midrule
50   &2028 &4680 &67.96 &134.23 \\ %
100  &2284 &3960 &52.79 &117.34 \\ %
200  &2540 &3735 &45.20 &111.68 \\ %
\bottomrule
\end{tabular}
\end{table}

\subsubsection{Participation costs}
\label{sec:results_participation_costs} 

We profiled and collected data on resource usage by participants of \system\ in the fully honest case of \S\ref{sec:results_perf_evaluation}. In this scenario, around $45,000$ new polls are created per day, and every user on average takes part in $\approx 5$ polls\footnote{See Appendix~\ref{app:cost_determining_eligibility_polls} for details.}. Similar to \blockene, the data usage in \system\ was observed to be 20MB/block committee and the cost of periodic \textsf{GetLedger}s (\S\ref{sec:blockene_basics}) was $21$MB/day. With an observed block latency of $\approx 72$ seconds, corresponding to 1200 blocks/day, each user on average participates in $\approx \frac{2000 \times 1200}{10^6} = 2.4$ block committees per day (\ie $40-60$MB/day total). Thus, the total network usage for a \citizen in our system is $70$MB/day on average -- only a slight increase over \blockene's $60$MB/day usage.

Additionally, to vote in a poll with ring size $100$, a mobile client spends ${\approx}120$ms (Table~\ref{tab:urs_sign_verify}) in computing the signature, and the size of this vote would be $\approx2.3$KB. Broadcasting this new vote to 200 \politicians\ would require $\approx 460$KB of network upload. On a $1$ MB/s network connection, the combined time to compute the signature \textit{and} upload it is less than one second, a reasonably low cost that users incur when reviewing an article.

Keeping the committee cost and system throughput similar to Blockene was possible due to our aggressive offloading techniques. With a constant cost of ~$70$ MB/day network, and an additional 460KB per new vote, we believe \system\ can be reasonably deployed at scale.

%% file: sections/7_discussion.tex
\section{Discussion}
\subsection{Usecases}
At its core, what \system\ provides is a framework to obtain anonymous responses to polls and efficiently commit it to a verifiable, distributed data store. In the permissioned usecase, where organizations are conducting opinion polls or surveys, this committed raw data itself is the final product - anyone can verifiably download the results of each poll.

In the permissionless usecase (\eg as a news aggregator), the end goal is to use this collected data to build a ranking system. In most online platforms (e.g. Reddit), each new post is published inside a community (``{subreddit}") organised around similar interests. The decision on what shows up on the front page of that community is made by a black-box ranking engine controlled by the platform. In \system, each of these communities would be a topic (\S\ref{sec:main_body_topics}) and each new piece of content would be a \textit{poll}, for which users are selected (\S\ref{sec:random_sampling}) to review. Since reviews are readily retrievable from the global state, a wide range of content-based~\cite{content-based} ranking/recommendation algorithms can be built as a ``front-end" on top of \system. Potentially, multiple such frontends could exist, with users deciding which one they wish to use, instead of the platform making that determination for them.

\subsection{\system\ Scale}
The next question we seek to answer is that of scale - given the evaluation we have seen, what kinds of communities can be realistically launched on \system\ based on real-world examples? To get a sense of the size (number of users) and the volume (number of new posts per day) of online communities we turn towards studies analysing usage patterns on existing platforms today. A recent meta-analysis~\cite{reddit-systematic-overeview} explicitly identified 415 active communities on Reddit where the total number of subscribers ranges between ten thousand and one million people, while another study~\cite{anatomy-of-reddit} estimates this number to be in the thousands. These communities span topics ranging from mental health~\cite{dreaddit-2019} and medicine~\cite{reddit-dermatology} to politics and current affairs~\cite{reddit-political-communities} among others. Any of these communities would be a suitable target for migration to \system, since our experimental results demonstrate scalability upto a million users. Even with topics that generate sudden influx of discourse online, such as the Russia-Ukraine war, studies estimate the number of new posts created to be around $~3,000$ on average per day on Reddit~\cite{russia-ukraine-reddit-dataset} and a peak of $~1,500$ on TikTok~\cite{tiktok-invasion-ukraine}. From the results section, we notice that \system\ supports up to $45,000$ new polls (\ie new pieces of content) per day, providing evidence that \system\ would be suitable even for topics that experience surges in popularity.

\subsection{User participation in \system}
\subsubsection{Permissionless usecase}
Unlike existing content ranking platforms (\eg Google News,  Reddit, YouTube), where users peruse (and provide feedback on) whatever content they want, \system's permissionless usecase calls upon users to provide reviews on specific entities they are randomly chosen for (\S\ref{sec:random_sampling}). While this does represent a change from the way these platforms function today, we believe that requiring users to volunteer occasionally is not very far removed from reality. Indeed, self-moderation is prevalent in most online communities~\cite{moderation-on-reddit-splotlight}, and previous studies have established how users of online platforms have strong intrinsic motivation to contribute to communities they are a part of~\cite{stackoverflow-motivation,wikipedia-intrinsic}; other studies have shown empirical evidence that gamification (\eg StackOverflow's badge system) stimulates volunteer contributions~\cite{cscw15-gamification-stackoverflow}, another strategy a \system\ deployment could employ. Additionally, \system\ (a) ensures that every user is selected to review new content relatively infrequently, and (b) is easily customisable to each user, through the use of topics (\S\ref{sec:main_body_topics}) where users subscribe to broad topics they are interested in (\eg history, biology, environment) and are only assigned review entities based on these interests; both these properties make it easier for users to actively participate.

\subsubsection{Permissioned usecase}
Existing opinion polling or survey systems can, on the other hand, be seamlessly replaced with \system, since employees/users being randomly asked to participate in opinion polls or surveys is the norm even today.

\color{black}

%% file: sections/8_related_work.tex
\section{Related Work}
\label{sec:related_work}

Protocols for content curation~\cite{Crites2020ReputableLC,May2014FilterF,kiayias_et_al:OASIcs:2020:11967} have studied how to aggregate opinions on various issues and ensure that users of the underlying system only see useful and relevant content. Typically, the magnitude of content and the constant influx requires some amount of crowdsourcing to obtain timely and user-judged opinions. Few works~\cite{TranNSDI09,Yu_sybil} also consider Sybil-resistance in addition to aggregation. In our work, $\system$ is also able to aggregate opinions to various polls in an efficient and scalable fashion, but unlike existing work in this space, $\system$ is both private and hijack-resistant, which prevents any group from obtaining an outsized influence on any poll, and also prevents linkability of votes to the real identity of the voter. 

e-Voting protocols like those in ~\cite{Liu2017AnEP} allow for privacy-preserving voting as well as aggregation of votes in a verifiable fashion without a trusted central authority. Boardroom voting protocols~\cite{kulyk14, Groth2004EfficientMP} are designed for a small number of voters who all have the same power - all of them vote as well as participate in tallying the votes. In a similar vein, Zhang et al.~\cite{DBLP:conf/ndss/ZhangOB19} develop a treasury system that selects a random subset of parties (weighted by stake similar to Algorand \cite{algorand}) every epoch to form a voting committee for proposal and provide techniques to aggregate votes by using unit vector encryption schemes based on ideas from \cite{one_many}. However, these schemes are designed for relatively low-frequency events (like elections, or funding decisions), while we are able to handle millions of votes and tens of thousands of polls per day at a significantly lower participation cost to the voters.

Kiayias et al. \cite{cryptoeprint:2022/760} develop a privacy preserving protocol for opinion aggregation to issues. They focus on continuous aggregation for any poll, where votes are aggregated in batches indirectly providing anonymity up to the batch size. In $\system$, we ensure that votes are anonymous individually, and do not have to rely on other voters to get a large enough anoynomity set. They also rely on voters to participate in the system which increases the voter workload beyond that of just voting. In addition, they do not propose an end-to-end system, and use general purpose ZK proofs to obtain verifiability and privacy, which are inefficient on a large scale.

Importantly, none of the existing literature, to the best of our knowledge, considers hijack resistance as a desirable criterion for voting/review systems. We define a quantitative notion of hijack resistance and demonstrate that \system\ achieves strong hijack resistance in a decentralized setting while also preserving anonymity.

%% file: appendix/1a-blind-sigs.tex
\section{Background on Blind Signatures}
\label{app:blind-sigs}

Blind signatures \cite{blind_sigs_bellare2003one, blind_sigs_chaum1982} enable users to obtain signatures on a message of their choice without revealing the message itself to the signer. The scheme consists of five algorithms:
\begin{itemize}
    \item $\mathsf{BlindSig.Setup}$
    \item $\mathsf{BlindSig.Blind}$
    \item $\mathsf{BlindSig.Sign}$
    \item $\mathsf{BlindSig.Unblind}$
    \item $\mathsf{BlindSig.Verify}$
\end{itemize}

The setup $\mathsf{BlindSig.Setup}$ is run by the signer to generate the public ($pk$) and secret ($sk$) keys of signature scheme. 
To get a signature on message $M$, the user first blinds the message $M$ using $\mathsf{BlindSig.Blind}$ with randomness $r$ and public key $pk$ to get blinded message $M'$. 
Signer runs $\mathsf{BlindSig.Sign}$ on $(M', sk)$ to produce $S'$ that is sent to the user. 
The user unblinds $S'$ using $(r, pk, M)$ to obtain a signature $S$ on $M$ that should verify under $pk$ with $\mathsf{BlindSig.Verify}$.

A blind signature scheme should satisfy the standard correctness guarantee that a faithfully generated signature obtained by following the above procedure is a valid signature on $M$ under $pk$. It needs to satisfy two security properties.

\begin{enumerate}[leftmargin=0.2in, itemsep=1pt]
    \item \textbf{Unforgeability:} A malicious user that interacts with the signer $k$ times cannot produce $k+1$ valid message-signature pairs, for any $k$.
    \item \textbf{Blindness:} A malicious signer that interacts with a user in $k$ sessions to produce $k$ signatures, cannot map any signature to the session which was used to produce it (better than at random) for any $k$.
\end{enumerate}

While our scheme is compatible with any blind signature scheme, our implementation uses RSA blind signatures %

\begin{figure}[htbp]
    \centering
    \includegraphics[width=\linewidth]{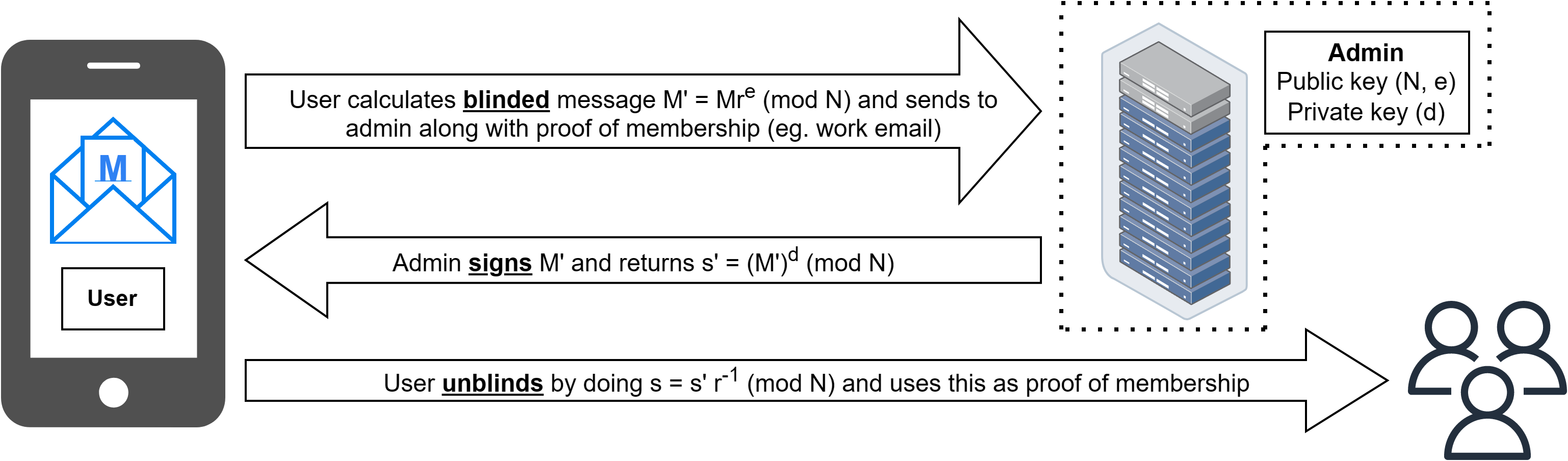}
    \caption{Overview of the blind signature protocol. Once the user unblinds the message received from the admin, they can send it to third parties as proof of membership.}
    \label{fig:rsa_blinding_overview}
\end{figure}

\subsection{ RSA Blinding Microbenchmarks}
\label{app:rsa_blinding_microbenchmarks}
We provide an estimate on the computational cost of RSA blind signatures by reporting runtimes from a publicly available JavaScript library \cite{blind-sigs}. We performed these experiments on a VM running Ubuntu 20.04 LTS, Node.js version v10.19.0 on a single core of a Intel(R) Xeon(R) E5-2673 v4 CPU (similar hardware configuration to that of a \politician). Table~\ref{tab:rsa_blinding_microbench} presents these results, where all numbers are the average and standard deviation over 10 runs. From Table~\ref{tab:rsa_blinding_microbench}, we notice that the two steps that take the most time are the (1) KeyGen, (2) Signing. In \system's permissioned setting (\S\ref{sec:rsa_blinding}), both of these are steps performed by the admin of the system. All other steps -- blinding, unblinding and verification -- are performed by every user of the system and are relatively cheap (taking only a few milliseconds of CPU time each). 

\begin{table*}[htpb]
\scriptsize
\centering
\caption{Time taken for each step in the blind signature process (single core).}
\begin{tabular}{cccccc}\toprule
Key Size (bits) & KeyGen (s)        & Blinding (ms) & Signing (ms)     & Unblinding (ms) & Verifying (ms) \\\midrule
2048            & $2.85 \pm 1.85$   & $7.19\pm8.50$ &$171.78\pm18.18$  & $11.65\pm12.69$ & $4.24\pm5.90$  \\
3072            & $22.49 \pm 25.93$ & $4.35\pm0.55$ &$515.39\pm19.00$  & $6.95\pm2.20$   & $4.28\pm4.88$  \\
4096            & $40.02 \pm 29.36$ & $7.17\pm0.84$ &$1226.39\pm45.38$ & $11.95\pm1.45$  & $4.61\pm0.47$   \\ \bottomrule
\end{tabular}
\label{tab:rsa_blinding_microbench}
\end{table*}

Now, we will also try to estimate the amount of network traffic this RSA blinding phase generates. From Figure~\ref{fig:rsa_blinding_overview}, we see that there are three exchanges of information -- in each of those exchanges, the amount of data transmitted is equal to the bitsize of $N$ (since the numbers being sent are $\bmod\; N$). For a bitsize of $2048$, the network compute a \textit{user} would expend in performing this protocol is $\frac{2048 \times 3}{8} = 768$ bytes. In a system with one million users, the total network usage by the admin server would be around $512~\mathsf{bytes} \times 10^6 = 512MB$. Similarly, the total time required by the admin server to perform this blinding operation one million times (assuming each signing operation takes $\approx 171$ms; Table~\ref{tab:rsa_blinding_microbench}) would be $\approx 47.5$ hours on a single core, or $\approx 3$ hours on an 16-core machine. We believe these costs are reasonable since (1) the amount of compute and network required on end user devices is minimal, and (2) this is a cost that has to be paid rarely (\eg once in a month). 

%% file: appendix/2b-urs_performance.tex
\section{$URS_m$ Protocol}
\label{app:urs_m_details}

\noindent For the \URS{} scheme in \cite{log_urs} with algorithms $(\mathsf{\URS.Setup}$, $\mathsf{\URS.KeyGen}$, $\mathsf{\URS.Sign}$, $\mathsf{\URS.Verify})$, our modified scheme $\mathsf{\URS_m}$ is as follows:

\begin{enumerate}[leftmargin=0.2in, itemsep=1pt]
    \item $\mathsf{\URS_m.Setup}$: Same as $\mathsf{\URS.Setup}$.
    
    \item $\mathsf{\URS_m.KeyGen}$: Same as $\mathsf{\URS.KeyGen}$.
    
    \item $\mathsf{\URS_m.Sign(M, R, sk_i)}$, for $M = (\pollID, \voteValue)$: Obtain signature $S = (\nu, \sigma)$ by running $\mathsf{\URS.Sign(\pollID, R, sk_i)}$, where $\nu = H(\pollID || R)^{sk_i}$.
    Let $\tau = H(M||R)^{sk_i}$. Produce a (non-interactive) proof of equality of discrete logarithms for $\tau, \nu$ with bases $H(M || R)$ and $H(\pollID || R)$, respectively, as $\pi$. Output the signature as $(S, \tau, \pi)$.
    
    \item $\mathsf{\URS_m.Verify(\voteValue, M, R, \sigma)}$: Run $\textsf{\URS.Verify}(\pollID, R, S)$ and if it accepts, run the $dlog\_eq$ verifier on $(\nu, \tau, \pi)$ to check equality of discrete logarithms between $\tau$ and $\nu$. 
    Output $1$ iff both the verifiers output $1$.
\end{enumerate}

The sigma protocol for equality of discrete logarithms $dlog\_eq$ is given in Fig.~\ref{fig:dlog_eq} and can be made non-interactive using the Fiat-Shamir heuristic~\cite{fiat1986prove}.

We sketch how we show correctness, anonymity, unforgeability and uniqueness according to the definitions in \cite{log_urs}. 

\begin{enumerate}[leftmargin=0.2in, itemsep=1pt]
    \item Correctness follows from the correctness of the \URS{} scheme  as well as that of the sigma protocol for $dlog\_eq$.
    
    \item Anonymity in the ring follows from the anonymity of the original \URS{} scheme as well as the zero-knowledge property of the $\Sigma$-protocol for  $dlog\_eq$.
    
    \item Unforgeability follows from the unforgeability property of the \URS{} scheme.

    \item Uniqueness holds with respect to $(\pollID||R)$ from the original \URS{} scheme. Since we produce two \textit{linked} tags $(\tau, \nu)$ where the first tag only depends on the poll ID and not the vote, we can show that any valid vote on a particular message by a given user must have a unique first tag. The  $dlog\_eq$ zero knowledge protocol binds together the PID and vote of the same user. This allows us to check for duplicate votes by comparing the first tag of two different signatures, and hence, ensuring a single vote per user. 
\end{enumerate}

\paragraph{Additional Optimization.} We implement an efficient batch algorithm for verifying multiple signatures for the same poll. This allows us to reduce the number of expensive group exponentiations and convert them to multiplications and scalar arithmetic. We provide more details in Appendix~\ref{app:batching_urs}. We also use the Ristretto group \cite{ristretto} to \emph{safely} implement groups of prime order as required by the protocol. Operations over this group are as fast as elliptic curve operations, while certain attacks are also prevented by avoiding elliptic curves with small cofactors, which have been known to introduce vulnerabilities when used as a drop-in replacement for prime-order groups. These allow us to improve significantly on the performance shown in~\cite{log_urs}. More details and microbenchmarks are summarized in Appendix \ref{app:batching_urs}. 

\begin{figure}[h]
    \centering
    \fbox{%
        \procedure{dlog\_eq: Proof of equality of discrete logarithms}{%
            \text{Claim}: \tau = g_2^x, \nu = g_3^x \\
            \quad 1.\; \text{Prover sends } m_1 = g_2^r, m_2 = g_3^r \in \mathbb{G} \text{ to the verifier.} \\
            \quad 2.\; \text{Verifer sends } h \sample \{0,1\}^\lambda. \\
            \quad 3.\; \text{Prover computes and sends } z = hx + r \text{ to the Verifier}. \\
            \quad 4.\; \text{Verifier accepts if } g_2^z = m_1 \tau^h \text{ and } g_3^z = m_2 \nu^h.
        }
    }
    \caption{Proof of equality of discrete logarithms}
    \label{fig:dlog_eq}
\end{figure}

\subsection{Performance of URS}
\label{app:batching_urs}

\subsubsection{Signature Size}
For a ring of size $N$, each proof consists of $5n+4$ group elements and $3n+2$ scalars, where $n = \left \lceil \log_2(N) \right \rceil$. In the Ristretto group, both scalars and group elements are encoded as $32$-byte numbers. Thus, the total memory required for each signature is $32 \times (8\left\lceil\log_2{N}\right\rceil + 6)$ bytes. Even for a key-ring with 16,000 keys in it, the signature itself would still be around 3.7kB. 

\subsubsection{Sign and Verify Times}
Table \ref{tab:urs_sign_verify} shows the time taken on a OnePlus Nord CE smartphone for signing and verifying one message in this URS system, for various sizes of keyrings.
\begin{table}[!htp]\centering
\caption{Time taken (\textit{ms}) for Sign and Verify for various ring sizes on a smartphone} \label{tab:urs_sign_verify}
\scriptsize
\begin{tabular}{lrrr}\toprule
N &Sign &Verify \\\midrule
128 &134.85 &26.00 \\
256 &299.95 &44.97 \\
512 &674.26 &81.99 \\
1024 &1489.06 &155.47 \\
2048 &3278.77 &299.89 \\
4096 &7147.05 &590.87 \\
\bottomrule
\end{tabular}
\end{table}

\subsection{Optimization: Batching and Verifying Multiple Signatures}
\label{app:batch_verification_details}

\noindent \S3 of \cite{log_urs} describes the Verify protocols for checking if a given signature is valid. We use the same notation they use. Let us say we have a key-ring with $N$ people in it, and $M$ signatures all signed on the same keyring. The following steps refer to the steps from the URS protocol in~\cite{log_urs}.

\begin{enumerate}
    \item \textbf{In Step 3} of the verify protocol, for a single signature, we need to check, for each $j = 1, \dots, n$, 
    \begin{equation}
     c_{l_j}^X c_{a_j} = g^{f_j} h^{z_{a_j}} \,,\; c_{l_j}^{X-f_j} c_{b_j} = h^{z_{b_j}}  
    \end{equation}
    For $M$ proofs, this would result in $n \cdot M$ iterations, if we were to verify each proof one by one. However, we can equivalently sample $2*n\cdot M$ random $r_k \in \mathbb{Z}_q$ and check these equations instead:
    \begin{equation}
        \begin{split}
            \prod_{k = 1}^{n\cdot M} \left(c_{l_{j,k}}^{X_k} c_{a_{j,k}}\right)^{r_k} & = \prod_{k = 1}^{n\cdot M} \left(g^{f_{j,k}} h^{z_{a_{j,k}}}\right)^{r_k} \\
            & = g^{\left(\sum_{k = 1}^{n\cdot M} f_{j,k} r_k\right)} \cdot h^{\left(\sum_{k = 1}^{n\cdot M} z_{a_{j,k}} r_k \right)}\label{eqn:batched_step3}
        \end{split}
    \end{equation}
    
    \begin{equation}
        \begin{split}
            \prod_{k = 1}^{n\cdot M} \left(c_{l_{j,k}}^{X_k - f_{j,k}} c_{b_{j,k}}\right)^{r_{k+nM}} & = \prod_{k = 1}^{n\cdot M} \left( h^{z_{b_{j,k}}}\right)^{r_{k+nM}} \\
            & = h^{\left(\sum_{k = 1}^{n\cdot M} z_{b_{j,k}} r_{k+nM} \right)}
        \end{split}        
    \end{equation}
    We see from the RHS of \ref{eqn:batched_step3} that we replaced $n \cdot M$ group exponentiations with $n \cdot M$ field additions, followed by just one exponentiation.
    
    \item \textbf{In Step 6}, we need to check that 
    \begin{equation}
    \prod_{i = 0}^{N-1} pk_i^{p_i(X)} \cdot \prod_{j = 0}^{n-1} c_{d_j}^{-X^j} = h^{z_d} \end{equation}
    This step would need to be done once per signature, or a total of $M$ times for $M$ signatures. 
    Instead, we can sample $M$ random $r_k \in \mathbb{Z}_q$ and instead check the batched equation:
    
    \begin{equation}
    \begin{split}
        \prod_{k = 1}^M & \left(h^{z_{d,k}}\right)^{r_k}  = \prod_{k = 1}^M \left( \prod_{i = 0}^{N-1} pk_i^{p_{i,k}(X_k)} \cdot \prod_{j = 0}^{n-1} c_{d_{j,k}}^{-X_k^j} \right)^{r_k} \\
        & = \prod_{i = 0}^{N-1} pk_i^{\left( \sum_{k = 1}^{M} p_{i,k}(X_k) r_k \right)} \cdot \prod_{k = 1}^M \left( \prod_{j = 0}^n \left( c_{d_{j,k}}^{-X_k^j} \right)^{r_k} \right)    
    \end{split}
    \end{equation}
    Similar to the previous batching, this saves group exponentiations on the first term and instead does field additions.
\end{enumerate}

Table \ref{tab:batch_verify_time} shows the per-signature time taken when batch verify multiple signatures in one go. We experiment with multiple batch sizes - ranging from 25\% of the ring size to 100\% of the ring size. These numbers are from running benchmarks on the same OnePlus Nord CE smartphone used to create Table \ref{tab:urs_sign_verify}. Comparing the time taken for batch verify with that of individual verification (from Table \ref{tab:urs_sign_verify}), we notice a $3.1\times$ reduction in time for rings of size $128$ and a $22\times$ reduction for $4096$ sized rings.

\begin{table}[!htp]\centering
\caption{Time taken (\textit{ms}) per signature during batch verification on a smartphone}\label{tab:batch_verify_time}
\scriptsize
\begin{tabular}{lrrrrr}\toprule
\multirow{2}{*}{N} &\multicolumn{4}{c}{Percentage of ring being verified} \\\cmidrule{2-5}
&25.00\% &50.00\% &75.00\% &100.00\% \\\midrule
128 &8.30 &7.89 &7.81 &7.75 \\
256 &9.38 &9.09 &9.01 &8.95 \\
512 &11.02 &10.76 &10.61 &10.57 \\
1024 &13.63 &14.92 &13.19 &13.17 \\
2048 &17.97 &17.85 &17.63 &17.70 \\
4096 &26.50 &26.41 &26.16 &26.25 \\
\bottomrule
\end{tabular}
\end{table}

%% file: appendix/3c-topics.tex
\section{Handling topics for polls and user interests}
\label{app:handling_topics}
In this section we discuss details around how users subscribe to polls (\S\ref{app:subscribing_topic}) and how the use of multiple topics impacts the hijacking cost (\S\ref{app:topics_hijack_cost}).

\subsection{Subscribing to topics}
\label{app:subscribing_topic}
\subsubsection{Permissioned Usecase}
A potential real-world deployment of the permissioned use case (\S\ref{subsec:intro_usecases}) could, for example, be used for conducting privacy-preserving surveys and polls within a large company. In this kind of scenario, it might be desirable if polls could be targeted at certain kinds of members (\eg a poll about market sentiment could be targeted to all sales staff, excluding everyone else). Here, the ``topic" that a user is subscribed to could be their role in the organization (eg. engineer, management, operations, etc). During the setup phase (\S\ref{sec:rsa_blinding}), for an organization with $N$ types of roles, the admin creates $N$ distinct \RSA\ key-pairs, one keypair for each topic. Every employee creates a $\URS_m$ keypair locally and performs the blind signature protocol with their respective admin \RSA\ key for their specific role, allowing them to derive a signature $\sigma$ on their public key.

\subsubsection{Public Usecase}
On the other hand, in the public use-case (disaggregated aggregators), topics are used primarily to group users by interest (\eg~history, sports, food, etc). Thus, users choose which category to subscribe to based on their interests, and could also be allowed to change their subscription periodically as their interests change. %

\subsection{Hijacking Cost}
\label{app:topics_hijack_cost}
Let us say there exists a topic $\categoryID_k$ with $\totalvoters_k$ subscribers in total. The hijacking cost \textit{for this topic} ($\pollbot_k$), thus, can be calculated using Lemma~\ref{lemma:hijacking_cost}, but replacing the total number of polls in the system $P$ with the number of polls in topic $t_k$. $\pollbot_k$ calculated this way can let us estimate what percentage of $\categoryID_k$'s subscribers need to be hijacked to achieve certain hijack goals ($\hijackedpolls_k, \hijackthreshold_k)$. 

\subsection{Allowing for user subscription changes}
\noindent An instance of \system\ that supports topics in the manner described in this section would need to add support for one more type of transaction, above and beyond the three basic transaction types defined in \S\ref{sec:trustrate-over-blockene}:
\begin{itemize}
    \item $\textsf{ModifySubscription}$: In the permissioned use case, modifying subscriptions is not allowed, and all such transactions are immediately discarded. In the permissionless use case, users are allowed to change their subscriptions - after verifying if the transaction and signature on it are valid, the updated subscription information is written to the global state. However, in this case, a subscription update is treated similarly to a new user registration, and that user is not allowed to participate in any new polls for the next $\seedwaitperiod$ blocks. 
\end{itemize}

%% file: appendix/4d-apathy.tex
\section{Handling User Apathy and Unavailability}
\label{app:dynamic_thresholding}
On a poll that requires a minimum of $\mathcal{R}$ votes, selecting only the top-$\mathcal{R}$ \VRFpoll s using Equation~\ref{eqn:vrf_ring_selection} might lead to polls not receiving enough responses if users are inactive or unavailable. A real-world application needs to be robust to the cyclic nature of user interest in a topic. We demonstrate one mechanism that \system\ could use to to increase the likelihood that we get a sufficient number of votes on polls ($\numvotesreq$). we introduce a parameter, $\dynamicthreshold{}_e$, that is updated every epoch (which could be configured to happen every $\approx1000$ blocks). During the \VRFpoll{} sortition for \pollP{}, the top $(\dynamicthreshold_e\cdot \numvotesreq)$ are selected instead of selecting just the top $\numvotesreq$. At the end of each epoch, \dynamicthreshold{} is updated. The new threshold ($\dynamicthreshold_{e+1}$) is calculated by keeping track of the following information during previous epoch $e$:

\begin{enumerate}[leftmargin=0.2in, itemsep=1pt]
    \item $v_{exp}$: For all polls whose voting windows start and end strictly within the epoch boundaries of epoch $e$, this is the total number of expected votes. 
    \begin{equation}
    \label{eqn:v_exp_calculation_new}
        v_{exp} \coloneqq \sum_{P \in \textsf{epoch}~e} \numvotesreqP{}
    \end{equation} 
    \item $v_{seen}$: For all whose voting windows start and end strictly between the epoch boundaries, $v_{seen}$ is the cumulative count of reviews that were submitted by eligible voters during the voting window for the poll. 
\end{enumerate}

\noindent At the end of the epoch, the new threshold $\dynamicthreshold_{e+1}$ for the next epoch is calculated by: \begin{equation}
    \label{eqn:threshold_update_new}
    \dynamicthreshold_{e+1} = \lambda \dynamicthreshold_{e} \cdot \frac{v_{seen}}{v_{exp}}
\end{equation} where $\lambda \geq 1$ is some constant.

\noindent One issue with Equations~\ref{eqn:v_exp_calculation_new}, \ref{eqn:threshold_update_new} arises when \numvotesreqP{} varies a lot between polls. To ensure that large polls (with high \numvotesreqP{}) are not disproportionately influencing threshold updates, we normalize polls' \numvotesreqP{} to a fixed value, $n'$: \begin{align}
    \label{eqn:normalized_vrf_update_new}
    v'_{exp} &\coloneqq \sum_{P \in \textsf{epoch}~e} n' \\
    v'_{seen} &\coloneqq \sum_{P \in \textsf{epoch}} \left(\frac{n'}{\numvotesreq}\right) \cdot \numvotesseenP \\
    \dynamicthreshold_{e+1} &= \lambda\dynamicthreshold_e \cdot \frac{v'_{seen}}{v'_{exp}}
\end{align} 
In the rest of this paper we stick to the simpler definitions of threshold updates (in Equations~\ref{eqn:v_exp_calculation_new},  \ref{eqn:threshold_update_new}) in lieu of these normalized forms for simplicity.\footnote{One problem with this dynamic system is that of \textit{permanently} inactive subscribers. If someone subscribes to a topic, and then just uninstalls the app or never participates in the system again, the thresholding system will still ensure that enough actual votes are recorded on each poll. However, this comes at an additional cost of a bigger ring size which might translate to longer signatures and higher sign/verify times (Appendix~\ref{app:batching_urs}). This, however, can be easily solved with a \textit{resubscription mandate}: every user must resubscribe to topics after a certain period of time (\eg a month).}

Notice that while calculating hijacking cost in Lemma~\ref{lemma:hijacking_cost}, we assumed that all people chosen to vote actually submit a vote - in particular, all honest people always vote when they are selected. This is not indicative of a real system, where users have certain periods of time where they do not vote even if selected to vote. 

To account for this, we consider a model where at most a fraction $\apathyfrac$ of the honest voters chosen in any poll do not vote, or are apathetic. This affects the hijacking cost, since bots/malicious users could potentially influence poll outcomes when this fraction $\apathyfrac$ of honest users do not vote. The worst-case effect of this apathy on the hijacking cost is shown in    Lemma~\ref{lemma:hijacking_cost_apathy}. %

%% file: appendix/5e-trustrateoverblockene.tex
\section{\NoCaseChange{Building \system\  over Blockene - Details}
}\label{appendix:trustrateoverblockene}
\begin{table}[!htp]\centering
\caption{Glossary of \system\ elements}
\label{tab:glossary}
\scriptsize
\begin{tabular}{ p{0.18\linewidth} m{0.1\linewidth} p{0.1\linewidth} p{0.47\linewidth} }\toprule
Attribute & Notation & Size (B) & Explanation \\\midrule

\multicolumn{3}{l}{\textbf{Common Attributes}} \\
SHA256 & $H(...)$ & \centering 32 & SHA256 hash of content in brackets \\
Block num. & $B_i$ & \centering 8 & Block $i$. \\

\\ \multicolumn{3}{l}{\textbf{Poll Attributes}} \\
Poll ID & \pollID & \centering 8 & Unique identifier for each poll.\\
Topic ID & $\categoryID_j$ & \centering 4 & Refers to the $j^{th}$ topic\\
Poll Data & \pollInfo & \centering 256 & The question being asked in the poll. \\
Req. votes & \numvotesreq & \centering 4 & Number of votes required on a poll \\
Seen votes & \numvotesseen & \centering 4 & Number of votes received on a poll so far \\ 
Vote Window & \votingWindow & \centering 4 & Number of blocks for which  poll is open for voting. \\

\\ \multicolumn{3}{l}{\textbf{Vote Attributes}} \\
Vote value & \voteValue{} & \centering 257 & 1B rating field (eg. 5-star scale) + 256B free-form text component (eg. review) of size 256-bytes \\
$\URS_m$ tag &  $\nu$ & \centering 32 & $\URS_m$ tags ensure uniqueness of votes \\ 

\\ \multicolumn{3}{l}{\textbf{Dynamic Thresholding}} \\
Votes exp. & $v_{exp}$ & \centering 4 & $\sum \numvotesreq$ of all polls in a certain topic, in current epoch \\
Votes seen & $v_{seen}$ & \centering 4 & Total votes received across all polls in a certain topic, in current epoch \\
\bottomrule
\end{tabular}
\end{table}

As alluded to earlier, our polling system is built as an application over \blockene, a blockchain system. A block committee in \blockene\ verifies transactions of our polling system and ensures that required data in these transactions (``messages'') is committed to the \textit{global state} (a Merkle tree). In this appendix, we explain:
 \begin{itemize}
    \item \textbf{Transaction syntax} All interactions with \system\ (i.e. registration, creating a poll, creating a vote, and subscription change) are done by users proposing a new \textit{transaction} for the action they wish to perform. The syntax for all these transactions is defined in Appendix~\ref{sec:txn_syntax}.
    \item \textbf{Global State Syntax} Appendix~\ref{sec:gs_syntax} describes the syntax in which data is stored in the global state, and explains how this syntax makes it efficiently queryable. 
    \item \textbf{Rules to validate transactions} Appendix~\ref{sec:txn_valid_rules} explains the rules that block committees use to determine if a transaction is (logically) valid or not. Invalid transactions are discarded and result in no changes to the global state.
    \item \textbf{Updating the global state} Appendix~\ref{sec:app_gs_update} describes the format in which data is stored in the global state, and the rules followed by block committees when writing to the global state.
\end{itemize}

\noindent Table~\ref{tab:glossary} provides a list of symbols used in this section, what they mean, and how much storage each element requires (in bytes). Throughout this section, we discuss the system design for \system\ with all the optional additions discussed in Appendices \ref{app:handling_topics} and \ref{app:dynamic_thresholding}. 

\subsection{Transaction Syntax}
\label{sec:txn_syntax}
\begin{table*}[htbp]
\centering
\caption{Transaction Namespacing \& Syntax. Vote transactions have signatures of varying length based on the size of the $\URS_m$ ring size ($|R|$).}

\label{tab:txn_namespacing}
\scriptsize
\begin{tabular}{llllrr}\toprule
    \textbf{Operation} &Key Format &Value Format &Message Size & Signature & Signature Size\\\midrule 
    \textsf{RegisterVoter} & "R" & $\URS_{pk}$, $\categoryID_j$ &  1+36=37 & \eddsa{} & 64\\
    \textsf{CreatePoll} & "P", \pollID & $\categoryID_j$, \numvotesreq, \votingWindow, \pollInfo & 9+268=277 & $\eddsa$ & 64\\
    \textsf{CreateVote} & "V", \pollID & \voteValue & 9+257=263 & $\URS_m$ & $32 \times (8\left\lceil\log_2{|R|}\right\rceil + 6)$\\
    \textsf{ModifySubscription} & "C" & $\categoryID_j$ &1+4=5&\eddsa & 64 \\
    \bottomrule
\end{tabular}
\end{table*}

Table~\ref{tab:txn_namespacing} lists the transaction  formats of our application and the signature scheme required to sign these transactions when they are proposed. \begin{itemize}
    \item \textsf{RegisterVoter} transactions require \eddsa{} signatures (for the permissionless system) or \textsf{RSA} signatures (for the permissioned system).

\end{itemize}

\subsection{Global State Syntax}
\label{sec:gs_syntax}

\begin{table*}[ht]\centering
\caption{Global State Syntax}\label{tab:gs_namespacing}
\scriptsize
\begin{tabular}{llllr}\toprule
\textbf{\#} & \textbf{Type} &Key Format &Value Format &KV Size \\\midrule
1&\textsf{Poll} & "P", \pollID & $B_{n}$, $\categoryID_j$, \numvotesreq, \numvotesseen, \votingWindow{}, $H(R)$ &9+56=65\\
2&\textsf{PollData} & "I", \pollID & \pollInfo &9+256=265\\
3&\textsf{Vote} &"V", \pollID, $k$ & \voteValue & 10+257=267\\
4&\textsf{VoteTag} &"T", \pollID, $\nu$ & \textsf{True} & 10+1=11\\
5&\textsf{Thresholds} &"W", $\categoryID_j$ & $W_j$ & 5+32=37\\
6&\textsf{ScalingFactor} &"U", $\categoryID_j$ & $v_{exp}, v_{seen}$ & 5+8=10\\
7&\textsf{BlockwisePolls} &"B", $B_i$, $\categoryID_j$ & $H(\pollID_1 || \pollID_2 \ldots)$ &13+32=45\\
\bottomrule
\end{tabular}
\end{table*}

In each block, the block committee processes the transactions and updates the global state based on these transactions. Table~\ref{tab:gs_namespacing} shows how data is stored in the Global State, as well as the size of each of these data entries in bytes. We now explain each of the entries in Table~\ref{tab:gs_namespacing}: 

\begin{itemize}[leftmargin=0.1in, itemsep=0pt]
    \item \textsf{Poll} and \textsf{PollData} together store information on new polls. 
\begin{itemize}[leftmargin=0.1in, itemsep=0pt]
        \item \textsf{Poll} contains essential information or meta data about the poll (i.e. information that is needed to verify a new vote). This includes the block number in which the poll was committed ($B_n$), number of votes seen so far (\numvotesseen{}), the \votingWindow{} for the poll and a succinct representation for eligible voters $H(R)$ where $R$ is the ring to be used in ring signatures during voting\footnote{See Section~\ref{sec:results_participation_costs} for a description on how users identify if they are part of the poll}. When a poll is committed, $R$ is unknown and only gets updated after $\seedwaitperiod$ number of blocks as discussed in Section~\ref{sec:app_gs_update}.
        
        \item \textsf{PollData} contains the text of the question, along with any other information provided by the \creator{} of the poll. This data field, stored as 256-byte free-form text field, and is not required to verify a new vote transaction (the \pollID{} of a poll is sufficient to identify a new poll). We store this data separately so that any party that wishes to merely verify a vote does not have to download all this other information. 
    \end{itemize} 

    \item \textsf{Vote} and \textsf{VoteTag} together represent votes on a poll.
\begin{itemize}[leftmargin=0.1in, itemsep=0pt]
    \item \textsf{Vote} maps \pollID\ and an index $k$ to corresponding vote $\voteValue$ containing the value of the vote (\voteReview) and an optional comment (\voteComment). 
     \item \textsf{VoteTag} stores the tag $\nu$ of this vote along with \pollID. This will be used to ensure that each user can vote only once.

\end{itemize}

    \item \textsf{Thresholds} contains the value of the current thresholds, while \textsf{ScalingFactor} contains information required to calculate the next thresholds after the current epoch is complete. 

    \item \textsf{BlockwisePolls} contains the hash of a sorted list of \pollID{}s committed in a certain block $B_{i}$ in a topic $\categoryID_j$.

\end{itemize}

\subsection{Transaction Validation Rules}
\label{sec:txn_valid_rules}
This section defines the logical checks each block committee performs before accepting a vote or poll transaction. Invalid transactions are immediately rejected and discarded. $B_i$ is the current block number.

\begin{itemize}[leftmargin=0.1in, itemsep=0pt]
    \item For \textsf{RegisterVoter} transactions:
    \begin{itemize}[leftmargin=0.1in, itemsep=0pt]
        \item $\URS_{pk}$ is a new and well-formed public key for URS.
        \item \categoryID\ is a one of the previously defined categories.
    \end{itemize}
    
    \item For \textsf{CreatePoll} transactions:
    \begin{itemize}[leftmargin=0.1in, itemsep=0pt]
        \item \pollID{} is distinct from all previous poll IDs. 
        \item $\categoryID_j$ exists.
        \item \numvotesreq{} is a positive integer smaller than $|\audience{}|$, i.e., number of registered users that have subscribed to $\categoryID_j$.
    \end{itemize}
    
    \item For \textsf{CreateVote} transactions: 
    \begin{itemize}[leftmargin=0.1in, itemsep=0pt]
        \item Ensure that voting on the poll has begun and that the voting window for the poll is not over. $$B_p + \seedwaitperiod < B_i \leq B_p + \seedwaitperiod + \votingWindow$$ where $B_p$ is the block number in which $\pollID$ was committed.
        \item Let $R$ be the ring for the unique ring signature scheme, i.e., the set of eligible voters for this poll. Check that the signature is valid signature under $\URS_m$ and $R$.
        \item The tag $\nu$ of the \URS{} signature is not repeated in any of the previous votes on the same poll, i.e., this is the only vote by the user on this poll.
    \end{itemize}

As we discuss next in Section~\ref{sec:app_gs_update}, the information required to verify a new vote - $B_p$, $R$, and previously seen tags - can be read from the global state corresponding to \pollID. 
\end{itemize}

\subsection{Updating the Global State}
\label{sec:app_gs_update}
A block committee processes a pool of transactions and commits to a consistent block consisting of a sequence of valid transactions. For each of the transactions, it runs the transaction validation checks discussed in Section~\ref{sec:txn_valid_rules} and updates the global state (GS) appropriately. The block committee for a block $B_i$ calculates the new global state $GS_{i+1}$ using the previous GS, $GS_i$, and the set of transactions in the block. We explain how the $GS$ is modified for each transaction type:

\begin{itemize}[leftmargin=0.1in, itemsep=0pt]
 \item For \textsf{CreatePoll} transactions: Add two entries to $GS$ with the structure as defined in rows 1 and 2 of Table~\ref{tab:gs_namespacing}.
 \item For \textsf{CreateVote} transactions: Given a vote on poll \pollID, increment  $\numvotesseen$  by 1 in key \textsf{Poll} correponding to \pollID. Also, add two entries in GS corresponding to this vote with the structure defined in rows 3 and 4 of Table~\ref{tab:gs_namespacing}.
 \item For \textsf{RegisterVoter} and \textsf{ModifySubscription} transactions: Each block in \blockene~\cite{blockene_osdi} possesses a data structure called the \textit{identity sub-block}, which contains information on new identities registered in that block. We use this data structure, with a minor modification to support storing the \categoryID\ ID the user is subscribed to as well. Since this is a separate, independent data structure, these two requests do not require any change to the global state.
\end{itemize}

\textbf{Additional task for block committee} Each round, the block committee also performs a lookup of new polls that are now open for voting, calculates the ring (i.e. list of eligible voters for the poll) and stores $H(R)$ to the global state. They populate the list of polls to do this for by read the GS for key \textsf{BlockwisePolls} corresponding to $B_i-\seedwaitperiod$ and all existing categories $\categoryID_j$ to obtain the corresponding hash of poll IDs committed $\seedwaitperiod$ blocks ago. It queries the politicians for a list of poll IDs and matches it against the committed hash. For each of these polls, say \pollID, it determines the list of eligible voters, i.e., the ring $R$ to be used during voting, using the public keys of all registered users and their subscription to categories. It uploads the ring $R$ to the politicians and updates the key \textsf{Poll} corresponding to \pollID\ with $H(R)$. This would be used in the future to verify votes as described in Section~\ref{sec:txn_valid_rules}.

%% file: appendix/6f-security_model_offloads.tex
\section{Security Model for Optimisations}
\label{app:optimisations_security_analysis}
Broadly, for each change we make to the underlying \blockene, we need to prove that the resulting system still satisfies the properties of \textit{liveness} and \textit{correctness}. 

Both ring calculation offload and $\URS_m$ vote verification offload use the notion of a \textit{safe sample} -- by querying multiple \politicians for the same piece of information, \citizens increase the likelihood that they are querying \textit{atleast} one honest node. We borrow the term of a "good" citizens from \blockene to refer to \citizens which include at least 1 honest \politician in their safe sample. The safe sample is chosen once per block and reused for all the sub-protocols; thus the number of ``bad" citizens (\citizens whose safe samples do not include even a single honest \politician) in \system is the same as those in Blockene. To argue the soundness of our protocols, it suffices to make the argument for a “good” Citizens alone.

\subsection{Ring Calculation Offload}
\label{app:security_ring_offload}

Algorithm~\ref{alg:ring_offload} describes the ring offload protocol. In this protocol, \citizens are attempting to verifiably learn the \textit{ring-hash}, a hash of list of eligible $\votersP_j$ for a certain new poll $\pollP_j$. In the best case, when all \politicians respond with the same value, the total network cost of this protocol is $(32\cdot m \textrm{ bytes}) = 0.8$KB per poll, where $m=25$ is the safe sample size, and we assume the output of the \textsf{Hash} function is 256 bits (or 32-bytes). 

The possible malicious behaviours a \politician could use to attack this protocol are: (1) provide \citizens with the incorrect ring-hash, or (2) refuse to respond to requests for the ring. The first form of attack is an attack on the \textit{correctness} of our protocol while the latter is an attack on the \textit{liveness}. Lemma~\ref{lemma:ring_offload_soundness} discusses the soundness of Algorithm~\ref{alg:ring_offload}. To prevent a single malicious node from causing an indefinite stall of the block committee in this step, we set a timeout ($\mathcal{T}_h$) on the request based on a calculation of the maximum conceivable time it could take for a honest \politician to respond. This calculation of $\mathcal{T}_h$ has two parameters - $t_{\mathsf{hash}}$ and $n_{\mathsf{thread}}$. The first parameter is the average time taken by a \politician to perform one $\mathsf{Hash}$ computation\footnote{For SHA256 hashes on our \politician nodes, we observed that $10^6$ hashes take $\approx1$ second on a single core.} and the second is the \textit{minimum} number of parallel threads a \politician is expected to have for performing hash computations. Both of these are parameters that can be determined before the system starts via microbenchmarks and analysing the system configuration. In our experiments (\S\ref{sec:implementation_and_results_section}), we noticed that honest \politicians are able to complete all required hash computations for rings \textit{well} in advance of when they actually become required (after \textsf{GSRead}). %

\begin{lemma}
\label{lemma:ring_offload_soundness}
    After following the steps of Algorithm~\ref{alg:ring_offload}, a good \citizen that obtained correct values from the \textsf{GSRead} protocol is guaranteed to have correctly computed ring hashes for each new poll. Additionally, this protocol will not indefinitely stall (liveness).  
\end{lemma}
\begin{proof}
    First, let us consider the case of a \citizen that receives responses from all the \politicians in the safe sample ($\mathcal{S}$) and all responses are identical. Since this is a good \citizen, atleast one \politician in its' safe sample $\mathcal{S}$ is honest, and thus, the downloaded hash values were correctly computed (since we assume an honest \politician performs these calculations faithfully and correctly). For the second case, consider a scenario when a certain subset of \politicians are unresponsive even after the timeout $\mathcal{T}_h$ but all the responses that were received are identical. In this case, as long as the parameters for $\mathcal{T}_h$ are set conservatively, honest politicians should have had more than enough time to perform the computations and respond. Finally, consider the case of having received conflicting hashes for a certain poll. Let $\votersP$ be the correct ring for the poll, and let us say we received $\votersP_1$ from politician $S_1$ and $\votersP_2$ from politician $S_2$. Without loss of generality, let $S_1$ be honest, and hence, $\votersP_1 = \votersP$. For any user $u \in \votersP_i$, $i \ne 1$, either $u \in R$ or the \VRFpoll\ of $u$ is lower than \VRFpoll\ of all members in $\votersP$ (since politician $S_1$ would have correctly computed this as an honest node).  
\end{proof}

\begin{algorithm*} 
\caption{Offload ring calculation}\label{alg:ring_offload}
\begin{algorithmic}[1]
\Require A list of $\pollID$s ($\pollID_1, \ldots, \pollID_{k}$) for which we need to calculate the lists of eligible voters. The \audienceP\ and ring-size required for each poll ($\mathcal{R}$). $\mathsf{Hash}$, a predetermined cryptographic hash function, and $\mathsf{Sort}$, a predetermined function that sorts UUIDs of $\votersP_j$ in lexicographic order. $\mathcal{S} = (S_1, \ldots, S_m)$, the \textit{safe sample} of \politicians used for \textsf{GSRead}.
\Ensure A good citizen that obtained correct values from \textsf{GSRead} learns the correct ring-hash for each poll, ($\mathcal{H}_1, \ldots, \mathcal{H}_k$) respectively, where ${\mathcal{H}_j\equiv\mathsf{Hash(Sort(\votersP_j))} \mid j\in[1,k]}$.

\State Ask all \politicians in $\mathcal{S}$ for the values of $\mathcal{H}_1, \ldots, \mathcal{H}_k$. 
\State If all \politicians in $\mathcal{S}$ responded and provided the same values, assume these values to be correct, and \textbf{exit}
\State If some \politicians are unresponsive, wait for a duration of time $\mathcal{T}_h=\frac{\lvert\audienceP\rvert\cdot k\cdot t_{\mathsf{hash}}}{n_{\mathsf{thread}}}$ for them to reply.
\State If after waiting $\mathcal{T}_h$, any \politician is still unresponsive, ignore them and proceed with the subset of responses received.
\For{every $\mathcal{H}_j$ for which we have received conflicting values $\{\mathcal{H}_j', \mathcal{H}_j'', \mathcal{H}_j''' \ldots\}$ }
    \State Ask every \politician in $\mathcal{S}$ that responded with a value of $\mathcal{H}_j$ to provide the set of voters they computed for $\pollID_j$
    \State Calculate the union ($\mathcal{U}$) of all the lists received \Comment{size of this list is bounded by $(m \cdot \mathcal{R})$}
    \State Calculate the $\VRFpoll$ (\S\ref{sec:random_sampling}) for each of the UUIDs $\in \mathcal{U}$, shortlist the top-$\mathcal{R}$ $\VRFpoll$s and $\mathsf{Sort}$ them. This set is $\votersP_j$
    \State Calculate the hash of this selected set of top-$\mathcal{R}$ UUIDs, and set this to be the final, correct value of $\mathcal{H}_j$
    \For{politician $S \in \mathcal{S}$ that returned a differing value of $\mathcal{H}_j$ from what was just calculated}
        \State \textbf{blacklist} S
    \EndFor
\EndFor
\end{algorithmic}
\end{algorithm*}

\subsection{Vote Verification Offload}
\label{app:security_vote_verification}
 Algorithm~\ref{alg:vote_verification__offload} presents the workflow of our vote verification offload protocol.
\begin{lemma}
\label{lemma:vote_verif_soundness}
    After following the steps of Algorithm~\ref{alg:vote_verification__offload}, a good \citizen is guaranteed to have correctly computed the verification status of every vote in the block. Additionally, this protocol will not indefinitely stall (liveness).
\end{lemma}
Proofs for liveness and soundness can be outlined in methods similar to those discussed in the previous section. One key difference in this protocol is that identifying the correct value of $b$ in case of conflicting responses from two politicians is simpler since the hash of the ring for every poll is committed to the Global State. To resolve conflicts a \citizen node querys one politician, verifiably retrieving the ring from it, and then verifies the vote locally, allow it to identify which politician was telling the truth and which was not. An additional thing to note is that the timeout $\mathcal{T}_r$ depends on the parameter $t_{\mathsf{verif}}$, the time taken per signature verification. Unlike, $t_{\mathsf{hash}}$ which is a constant, $t_{\mathsf{verif}}$ depends on the size of the ring.

\begin{algorithm*}
\caption{Offload vote verification}\label{alg:vote_verification__offload}
\begin{algorithmic}[1]
\Require A list of votes $V = [V_1, V_2, \ldots, V_k]$ on polls $P = [P_1, P_2, \ldots, P_j]$. $\mathcal{S} = (S_1, \ldots, S_m)$, the \textit{safe sample} of \politicians used for \textsf{GSRead}.
\Ensure A good citizen learns the correctly computed value of $B = [b_1, b_2, \ldots, b_k]$, a vector of boolean values, indicating whether each vote's signature is valid or invalid. 
\State Ask all \politicians in $\mathcal{S}$ for $B$. 
\State If all \politicians in $\mathcal{S}$ responded and provided the same vector, assume these values to be correct, and \textbf{exit}
\State If some \politicians are unresponsive, wait for a duration of time $\mathcal{T}_r=\frac{k \cdot t_{\mathsf{verif}}}{n_{\mathsf{thread}}}$ for them to reply.
\State If after waiting $\mathcal{T}_r$, any \politician is still unresponsive, ignore them and proceed with the subset of responses received.
\For{every $b_j$ for which we have received conflicting values $\{b_j', b_j'', b_j''' \ldots\}$ }
    \State Query random politician for the ring of $P_j$, along with Merkle challenge path to hash of the ring in the global state. 
    \State Use the downloaded ring to verify the vote locally, to get the correct value of $b_j$ 
    \For{politician $S \in \mathcal{S}$ that returned a differing value of $b_j$ from what was just calculated}
        \State \textbf{blacklist} S
    \EndFor
\EndFor
\end{algorithmic}
\end{algorithm*}

\subsection{Signature Validation Batching}
\label{app:security_batched_vote_verify}
As discussed in \S\ref{sec:opt_batchverify}, each pending transaction in \blockene is deterministically mapped to a specific \politician using a sortition function based on the hash of the transaction and the block number. 
Let us call this the \textbf{transaction mapping function}, $f_{\textsf{map}}$. 
After this mapping, the selected \politicians need to decide which subset of transactions to commit to, and \blockene does this randomly. 
Let us call this the \textbf{transaction selection function}, $f_{\mathsf{select}}$. 
This selection is needed because the size of the pending transactions could be far larger than the maximum number of transactions that \politicians can commit to in a single round.  
Note that malicious nodes might not respect or follow $f_{\textsf{select}}$ (\eg by dropping transactions), but all honest nodes must. Malicious nodes could also choose to select transactions that they were not mapped to, violating $f_{\textsf{map}}$, but this is easily detectable and blacklistable.

\blockene's security model guarantees that atleast $0.65$ fraction of blocks committed are non-empty (Theorem 2 of \cite{blockene_osdi}), and that honest \politicians select transactions to include in \txpools\ at random. 
Using these two facts, the authors of \blockene argue that every new transaction will \textit{eventually} be committed. 
We name the policies used in this default \blockene system to be $f_{\mathsf{map}} = \mathsf{DET\_TX\_HASHMAP}$ and $f_{\mathsf{select}} = \mathsf{RANDOM\_TX}$; the mapping function deterministically maps transactions to random \politicians using a hash, while the selection function selects transactions at random.

In \system, we suggest changing $f_{\textsf{map}}$ to a \textit{transaction group} based system ($\mathsf{DET\_GROUP\_HASHMAP}$), where an entire group of pending transactions is deterministically and randomly mapped to a specific \politician each round. 
Let us start by defining a couple of terms.
Let $T = \{t_1, t_2, \ldots, t_C\}$ be a list of pending transactions. 
Let us assume that each transaction $t_i$ is a member of \textit{one} transaction group $\mathcal{G}_j$ and $\{\mathcal{G}_1, \mathcal{G}_2, \ldots, \mathcal{G}_g\}$ are the set of all transaction groups ($g \leq c$). 
The simplest policy that can be chosen for $f_{\mathsf{select}}$ is randomness, \ie $\mathsf{RANDOM\_GROUP}$, which directs honest \politicians to randomly select transaction groups, and include all transactions in those groups in their commitment.

\begin{lemma}[Fairness of \textsf{RANDOM\_GROUP}]
\label{lemma:fairness_random_group}
    In a \blockene instance that uses an $f_{\textsf{map}} = \mathsf{DET\_GROUP\_HASHMAP}$ and $f_{\mathsf{select}} = \mathsf{RANDOM\_GROUP}$, that randomly maps each transaction group to a \politician, and then uses a random-group selection policy for $f_{\mathsf{select}}$, every transaction is eventually committed.  
\end{lemma}
\begin{proof}
    Since every honest \politician selects transaction groups at random, an arbitrary transaction group $\mathcal{G}_j$ is mapped to an honest politician with probability $0.2$ in each round; \ie over a large number of blocks, every transaction group is eventually selected to be in a \txpool. 
    Once \txpool\ commitments are made, there is a $0.65$ probability that a non-empty block is committed, meaning that atleast a subset of transactions in the \txpools\ are committed. 
    Using a similar line of argument as Lemma~14 of \cite{blockene_osdi}, we argue that every \textit{transaction group} $\mathcal{G}_j$ will eventually be committed. 
    Since, for every transaction $t_i$ there exists $j$ such that $t_i \in \mathcal{G}_j$, every transaction will eventually be committed.   
\end{proof}

It is possible, however, to do better than this. Consider a \blockene instance, where $f_{\textsf{map}} = \mathsf{DET\_TX\_HASHMAP}$ but $f_{\mathsf{select}} = \mathsf{OLDEST}$. The \textsf{OLDEST} policy directs honest nodes to sort transactions based on the timestamp at which they were received, and select the ones received longest ago (FIFO).  
\begin{lemma}[Fairness of \textsf{OLDEST}]
    In an instance of \blockene using $f_{\textsf{map}} = \mathsf{DET\_TX\_HASHMAP}$ and $f_{\mathsf{select}} = \mathsf{OLDEST}$, every transaction eventually gets committed 
\end{lemma}
\begin{proof}
    Let $t_o$ be the transaction with the \textit{oldest} received timestamp across all honest politicians $\mathcal{S} = \{S_1, S_2 \ldots S_h\}$. 
    The probability that this transaction $t_o$ is mapped to one of the honest nodes in $\mathcal{S}$ after $f_\textsf{map}$ is 0.2.
    If this transaction $t_o$ is mapped to any of the nodes in $\mathcal{S}$, it is guaranteed to be a part of \txpools\ because of the \textsf{OLDEST} policy for selection. 
    
    Using a line of argument similar to that of Lemma~\ref{lemma:fairness_random_group}, $t_o$ will eventually be committed since there is a $0.65$ probability of a non-empty block commit every time $t_o$ is present in the $\txpools$. 
    Once $t_o$ is committed, a new transaction $t_o'$ will now be the oldest transaction remaining, and using similar arguments, that, also will be committed. 
    Since every transaction eventually becomes the oldest remaining transaction in the pending pools of honest politicians, $\mathcal{S}$, every transaction is eventually committed.
\end{proof}

It is clear to see that the \textsf{OLDEST} strategy is likelier to lead to better transaction latency when compared to the \textsf{RANDOM} strategy for $f_{\mathsf{select}}$ - in the latter, the oldest transaction in a certain honest \politician's pending transactions pool \textit{might} be selected to be in the $\txpool$, while in the former it is guaranteed to be selected. 
In \system, recall that every poll \pollP\ has a voting window ($\votingWindow$) - a certain number of blocks before which votes must be committed in a block. 
In this scenario, it is possible to construct a selection function that selects the votes that are closest to the voting window expiring ($f_{\mathsf{select}} = \mathsf{DEADLINE}$). 
This, along with $f_{\textsf{map}} = \mathsf{RANDOM\_GROUP}$ could be used to form a uniquely tailored $\txpool$ policy that minimises CPU-cycles required through enabling the use of batching, while also making it likelier that votes don't miss their deadline through honest \politicians prioritising vote transactions that are closer to the deadline.

\subsection{Determining eligibility in polls}
\label{app:security_determining_eligible_polls}
\S\ref{sec:opt_eligibility_in_polls} describes the method used by members of \system\ to fetch the list of new polls they are eligible to vote on. Broadly, users ask \textit{all} Politicians to provide a list of all new polls they are eligible to vote on. Unlike the other protocols (\ie the ring and vote verification offloads), we notice that this protocol requires users to query \textbf{all} \politicians, instead of just a safe sample. The reason for this is because we have no margin for error here -- we have to \textit{ensure} that users come to know about all polls they are eligible for; a safe sample that contains zero honest nodes, in this scenario, would be able to drop polls that a user is eligible for. 

\subsubsection{Cost of this protocol} 
\label{app:cost_determining_eligibility_polls} 
From \S\ref{sec:results_perf_evaluation}, we see that throughput of votes in the all-honest scenario is 52.79 votes per second, or ${\approx}0.52$ polls/second in the steady state, which is around ${\approx}45,000$ polls per day. In a system with $10^6$ users, and with $100$ people being selected for each poll, it is expected that every user is selected for $\approx\frac{45,000 \times 100}{10^6} = 4.5$ polls/day on average. Let us assume that the user's app only checks for new polls once per day: \begin{itemize}
    \item First, the app queries all 200 \politicians for new polls and receives five new poll IDs from each. Total data downloaded: $5 \times 200 \times 8 \textrm{bytes} = 8$KB
    \item Next, for each poll, the app queries a randomly chosen politician for the ring of the poll (100 UUIDs, each $8$ bytes), along with a Merkle challenge path to the hash of the ring. The cost of downloading the rings is $5 * 100 * 8$ bytes = $4$KB.  Each challenge path is $300$ bytes (10-byte hashes, 30 levels deep), so the total network ingress required for this is 1.5KB for the challenge paths and $160$ bytes for the hash values themselves. 
\end{itemize} Summing all this up results in a total of ${\approx }13.6$ KB/day, a negligible cost for modern smartphones. 

%% file: appendix/7g-proofs.tex
\section{Proofs for Lemmas}
\label{app:lemma_proofs}
\begin{lemma}[Waiting period]
\label{thm:waiting_period}
    Assuming all new users are forced to wait for $B_{wait} = 38$ blocks before being allowed to participate in \system\, no adversary can predict inputs to the \VRFpoll{}s we use and violate the pseudorandomness of the output, except for negligible probability $\epsilon = 2^{-30}$.
\end{lemma}
\begin{proof}
    One input to our \VRFpoll{} is the $\mathsf{seed}$. Similar to \textit{Algorand}~\cite{algorand}, we define the seed for proposer $u$ as $\mathsf{seed}_r~\coloneqq VRF_{sk_u}(\mathsf{seed}_{r-1})$. Set the proportion of honest users, $h = 0.75$, since we assume the maximum corruption/malicious fraction in \blockene\ \citizen\ s~\cite{blockene_osdi} to be 25\%. Using arguments similar to that of Theorem 1 of \cite{algorand}, we have: \begin{align*}
        f(k) &\leq \left(\dfrac{(1-h)(1+h)}{h}\right)^{k-1} (1-h) \\
        &\approx 0.25 \cdot (0.583)^{k-1}
    \end{align*} 
    where $f(k)$ is the probability that $k$ continuous block proposers are malicious. To get $f(k) < 2^{-30}$, we can choose $k > 37.0079$, i.e $B_{wait} = 38$.
\end{proof}
\noindent The properties that our polling systems (for both our use cases) satisfies are:

\begin{lemma}[Uniqueness of votes]
\label{lemma:uniqueness_of_vote}
    No two votes $v_1 \neq v_2$ by the \textit{same} user and on the \textit{same} poll can be accepted by \textsf{ValidateVote} except with negligible probability.
\end{lemma}
\begin{proof}
    This follows directly from the uniqueness property of the $URS_m$ scheme that we use. Since \textsf{ValidateVote} checks that the tag $\tau$ in the transaction signature is unique and is not present in any previously seen vote on the same poll, the $URS_m$ scheme guarantees that the tag $\tau$ is a deterministic function of only the user's secret key, the \textsf{poll\_id} and the ring for the poll. Since this is unique for each user for a poll, no user can vote more than once and have it accepted by \textsf{ValidateVote}. 
\end{proof}

\subsection{Hijacking Cost Theorems}
\begin{lemma}
\label{lemma:hijacking_cost}
Let $0< \hijackthreshold, \hijackedpolls, \epsilon < 1$ be constants as defined in \S\ref{sec:def-pollng-system-properties}. \system\ is a review system with hijacking cost
$$\pollbot \coloneqq \frac{\hijackthreshold}{1 + g(\hijackthreshold \numvotesreq, \frac{\hijackedpolls}{1+g(\hijackedpolls \numberofpolls, \epsilon)})}$$ where $\numberofpolls$ is the number of entities reviewed in the system over some time period $T$ and $\numvotesreq$ is the number of unique reviews solicited for each poll, and  $$g(\alpha, \beta) = \frac{-3\ln \beta + \sqrt{(\ln \beta)^2 - 8\alpha \ln \beta}}{2\alpha + 2\ln \beta}$$ if every honest user selected to participate does indeed cast a vote.
\end{lemma}

\begin{proof}
    To simplify notation, define $g(\alpha,\beta)$ as the solution $x$ to $e^{-\frac{x^2 \alpha}{(1+x)(2+x)}} = \beta$. Explicitly, 
    $$g(\alpha, \beta) = \frac{-3\ln \beta + \sqrt{(\ln \beta)^2 - 8\alpha \ln \beta}}{2\alpha + 2\ln \beta}$$
    Since by Def.~\ref{defn:hijack_cost} we want more than $\hijackedpolls$ fraction of polls to be hijacked with probability at most $\epsilon$, the expected number of hijacked polls must be at most $E \coloneqq \frac{\hijackedpolls \numberofpolls}{1 + g(\hijackedpolls,\epsilon)}$. Notice that this follows from the Chernoff bound, since
    \begin{align*}
        Pr[\text{No. of hijacked polls } \geq \hijackedpolls \numberofpolls] &= Pr[\text{No. of hijacked polls } \\
        &\qquad \geq E \cdot (1 + g(\hijackedpolls \numberofpolls, \epsilon))] \\
        &\leq e^{-\frac{g(\hijackedpolls \numberofpolls, \epsilon)^2 E}{2+g(\hijackedpolls \numberofpolls, \epsilon)}} \\
        &= e^{-\frac{g(\hijackedpolls \numberofpolls, \epsilon)^2 \hijackedpolls \numberofpolls}{(1+g(\hijackedpolls \numberofpolls, \epsilon))(2+g(\hijackedpolls \numberofpolls, \epsilon))}} \\
        &= \epsilon
    \end{align*}
    This implies that each poll can be hijacked with probability at most $E/\numberofpolls$. 
        
    Now, the expected number of hijacked users selected for any poll would be $F = \frac{\hijackthreshold \numvotesreq}{1 + g(\hijackthreshold \numvotesreq, E)}$. Let the number (random variable) of hijacked users in a poll be $\mathcal{M}$. Using the Chernoff bound, we get
    \begin{align*}
        Pr[\mathcal{M} \geq \hijackthreshold \numvotesreq] &= Pr[\mathcal{M} \geq F \cdot (1 + g(\hijackthreshold \numvotesreq, E))] \\
        &\leq e^{-\frac{g(\hijackthreshold \numvotesreq, E)^2 F}{2+g(\hijackthreshold \numvotesreq, E)}} \\
        &= E/ \numberofpolls
    \end{align*}
    where we have the expected number of hijacked voters on any poll to be $E = \frac{\hijackthreshold \numvotesreq}{1 + g(\hijackthreshold \numvotesreq,\; E)}$. \\
    
    \noindent $\system$ is designed (using the properties of \VRFpoll) such that the expected fraction of hijacked voters is the fraction of corrupted users in the entire population of voters. Hence, $\system$ is hijack resistant with  
    $$\pollbot \coloneqq \frac{\hijackthreshold}{1 + g(\hijackthreshold \numvotesreq, \frac{\hijackedpolls}{1+g(\hijackedpolls \numberofpolls, \epsilon)})}$$ assuming that every selected \textit{honest} reviewer ends up casting a vote. Appendix~\ref{app:dynamic_thresholding} and Lemma~\ref{lemma:hijacking_cost_apathy} analyse what happens if some selected honest voters decide not to participate.  
\end{proof}

\begin{lemma}
\label{lemma:hijacking_cost_apathy}
    Given an apathy fraction $\apathyfrac$ of honest users, the hijacking cost of $\system$ changes from $\gamma$ as defined in Lemma~\ref{lemma:hijacking_cost} to $\gamma' \coloneqq \frac{\gamma (\apathyfrac-1)}{\gamma \apathyfrac - 1}$.
\end{lemma}
\begin{proof}
    Let the hijacking cost of the system with apathy be $\gamma'$ (the fraction of malicious users in the population). In the worst case, a poll is such that only $(1-\apathyfrac)$ fraction of the chosen honest users actually vote, and all the chosen malicious users vote. Hence, for any poll, the expected fraction of hijacked votes would be
    $$\frac{\gamma'}{\gamma' + (1-\gamma')(1-\apathyfrac)} = \frac{\hijackthreshold}{1 + g(\hijackthreshold \numvotesreq, \frac{\hijackedpolls}{1+g(\hijackedpolls \numberofpolls, \epsilon)})} = \gamma$$ 
    The quantity on the RHS is the hijacking cost $\gamma$ we derived in the previous lemma $\gamma$ when $\apathyfrac = 0$, i.e., no honest user is apathetic.
    Hence, 
    $$\gamma' = \frac{\gamma (\apathyfrac-1)}{\gamma \apathyfrac - 1}$$
\end{proof}

\subsection{Proofs for fairness and anonymity}
\label{app:proof_fairness_anonymity}

\begin{lemma}[Fairness of \VRFpoll]
\label{lemma:fairness_of_vrf1}
    Let us say a poll \pollP{} is using $\VRFpoll$ random sampling (Equation~\ref{eqn:vrf_ring_selection}) to select \voters{} from the \audience{}. Out of the \audience{} for this poll, a fraction of $x$ people ($0 < x \leq 1)$ form a \textit{group}, $\mathcal{G}$. For any such group, there exists a finite $\delta_f$, for which the fraction of \voters{} from $\mathcal{G}$ in \votersP is bounded by $x \cdot (1 \pm \delta_{f})$ with probability greater than $1 - \epsilon$, where $\delta_f \coloneqq \sqrt{\frac{3}{xN}\ln{\frac{2}{\epsilon}}}$.
\end{lemma}
\begin{proof}

    Let $M=|\audienceP|$, $N=|\votersP|$ and suppose there are $P = xM$ members in the group. Since \VRFpoll{}'s output looks random, we expect proportional representation, \ie the expected number of people in the group that are also voters in $\votersP$ is $xN$. To bound the deviation from this expected value, we can use a Chernoff bound; for any $\delta$, let the number of group members in $\votersP$ be $\eta$. Then,
    \begin{align*}
        Pr[\lvert \eta - xN \rvert \geq \delta xN] \leq 2e^{-\frac{\delta^2 xN}{3}}
    \end{align*}
    Set $\delta_f \coloneqq \sqrt{\frac{3}{xN}\ln{\frac{2}{\epsilon}}}$ for small $0 < \epsilon < 1$. Then, notice that
    \begin{align*}
        Pr[\lvert \eta - xN \rvert \leq \delta_f xN] &\geq 1- 2e^{-\frac{\delta^2 xN}{3}} \\
        &= 1-2e^{\ln{\frac{\epsilon}{2}}} = 1- \epsilon
    \end{align*}
    
\end{proof}

\begin{lemma}
    \system\ satisfies fairness (Defn~\ref{defn:fairness}).
\end{lemma}
\begin{proof}
    This follows from the fact that the opinion committee is selected at random from the set of all eligible users (guaranteed by \VRFpoll{}) and from Lemma \ref{lemma:uniqueness_of_vote} (which ensures that every eligible user can only cast one vote per poll). For some small probability $\epsilon$, the guarantee holds with $\delta_f$ given by the Chernoff bound (Lemma~\ref{lemma:fairness_of_vrf1}) due to the random sampling.
\end{proof}

\begin{lemma}
\label{lemma:content_anonymity}
    \system\ satisfies content anonymity (Defn.~\ref{defn:contentanonymity})
\end{lemma}
\begin{proof}
    Broadly, content anonymity derives its' basis from the $\URS_m$ scheme (Appendix~\ref{app:urs_m_details}). In \system, the identities of users ($\psid$s) are their respective $\URS_m$ keys. The $\URS_m$ protocol guarantees that an adversary possessing \textbf{only} the public keys of users in a ring (along with some signature $S$ from a member of the ring) will not be able to link a signature to a specific user in the ring with better than random probability (Appendix~\ref{app:urs_m_details}). Even if we assume users in \system\ generate their $\URS_m$ keys locally, and securely store their private keys, we will still have to argue that other components of our system do not leak information that could be used to deanonymize users. After users generate their $\URS_m$ key, it is used \textit{only} in two different occasions in the system: (1) when they initially register for the system and publish their public key to everyone in the system, and (2) when they participate in polls by signing their votes with their private key. In the registration phase, a malicious \politician could, in principle, obtain a device fingerpint~\cite{device_fingerprint} of the device performing the $\mathsf{RegisterVoter}$ (\S\ref{sec:trustrate-over-blockene}) request, creating a database mapping device fingerprints to $URS_m$ public key. Later, when the same device broadcasts a new vote the malicious \politician could deanonymize the user using information from the aforementioned database. Even if the registration phase itself was secure, similar probabilistic techniques could be used to fingerprint devices across multiple votes.  However, as discussed in \S\ref{sec:system_anonymity_and_fairness}, such attacks can be mitigated through the use of other techniques (\eg onion routing \cite{onion_routing}) that can be composed with our system. Assuming that these attacks can be prevented through techniques that anonymize network communication, content anonymity holds. 
\end{proof}

\begin{lemma}
    \system\ satisfies membership anonymity (Defn \ref{defn:membershipanonymity}) in the permissioned setting.    
\end{lemma}
\begin{proof}
    \system's membership anonymity is based on blind signatures (Appendix~\ref{app:blind-sigs}). Similar to the discussion in Lemma \ref{lemma:content_anonymity}, we assume that device fingerprinting based attacks are not possible and examine details of the blind signature protocol to identify situations that could lead to users' true $\id$s being linked to their $\psid$s with better than random probability. One scenario that could result in compromised membership anonymity is a \textit{timing-based} attack. Recall that users participate in an exchange with the admin, following which they publish the output of that interaction as a transaction on Blockene. This setup phase, involving interactions with the admin server, lasts only for a limited amount of time (say, $\mathcal{T}_{\mathsf{setup}}$), and all users are expected to finish the protocol in that time period. An important detail here is that throughout this time $\mathcal{T}_{\mathsf{setup}}$, honest users do \textit{not} send in \textsf{RegisterVoter} transactions; they must wait until this period is complete before registering. As an extreme example, consider a situation where the first user ($u_{\mathsf{first}}$) that participates in the $\mathsf{RSA}$-blind exchange protocol with the admin server at the beginning of $\mathcal{T}_{\mathsf{setup}}$, immediately sends in a \textsf{RegisterVoter} transaction before the admin server responds to any other requests. Here, $u_{\mathsf{first}}$'s $\psid$ is linkable to their $\id$ with probability $1.0$, which breaks content anonymity since this probability is significantly higher than random.

    Another important detail in our blind signature setup phase is that of registering and deregistering users. In the organisational use case, an administrator would want out system to support: (1) registering a new user (\eg new hire) and (2) removing existing user (\eg laid off employees). For both these changes, our system requires a complete \textit{fresh} start; with the admin creating and publishing a new $\mathsf{RSA}$ signing public-key, and \textbf{all} users (continuing users as well as brand new users) in the system generating new $\psid$s and participating in the blind signature protocol. This is required for similar reasons as the previous case; allowing a small set of new users to register keys later would reduce the degree of unlinkability they enjoy. In practice, performing a fresh start for every single change in employment might be excessive, and admins could use some policy to refresh keys periodically (\eg monthly) or after, say, there is $\approx1\%$ churn in the set of employees. This change would delay when new users would be able to join the system, or allow ex-members of the organisation to continue voting for a small period of time even after they leave, but comes with the advantage of preserving membership anonymity for all registered $\psid$s.
\end{proof}

%% file: main.bbl
\begin{thebibliography}{10}

\bibitem{android_ndk}
Android native development kit.
\newblock \url{https://developer.android.com/ndk}.

\bibitem{blind_sigs_bellare2003one}
Mihir Bellare, Chanathip Namprempre, David Pointcheval, and Michael Semanko.
\newblock The one-more-rsa-inversion problems and the security of chaum's blind signature scheme.
\newblock {\em Journal of Cryptology}, 16(3), 2003.

\bibitem{ddh_assumption}
Dan Boneh.
\newblock The decision diffie-hellman problem.
\newblock In {\em International algorithmic number theory symposium}, pages 48--63. Springer, 1998.

\bibitem{reddit-dermatology}
Talayesa Buntinx-Krieg, Joseph Caravaglio, Renee Domozych, and Robert~P Dellavalle.
\newblock Dermatology on reddit: elucidating trends in dermatologic communications on the world wide web.
\newblock {\em Dermatology online journal}, 23(7), 2017.

\bibitem{ethereum}
Vitalik Buterin et~al.
\newblock A next-generation smart contract and decentralized application platform.
\newblock {\em white paper}, 3(37):2--1, 2014.

\bibitem{nat}
Carrier-grade {NAT} - {Wikipedia}.
\newblock \url{https://en.wikipedia.org/wiki/Carrier-grade_NAT}.

\bibitem{cscw15-gamification-stackoverflow}
Huseyin Cavusoglu, Zhuolun Li, and Ke-Wei Huang.
\newblock Can gamification motivate voluntary contributions? the case of stackoverflow q\&a community.
\newblock In {\em Proceedings of the 18th ACM Conference Companion on Computer Supported Cooperative Work \& Social Computing}, CSCW'15 Companion, page 171–174, New York, NY, USA, 2015. Association for Computing Machinery.

\bibitem{social-media-manipulation}
Ho-Chun~Herbert Chang, Emily Chen, Meiqing Zhang, Goran Muric, and Emilio Ferrara.
\newblock Social bots and social media manipulation in 2020: The year in review, 2021.

\bibitem{blind_sigs_chaum1982}
David Chaum.
\newblock Blind signatures for untraceable payments. advances in proceedings of crypto 82, d. chaum, rl rivest, \& at sherman, 1982.

\bibitem{NTL}
A library for doing number theory.
\newblock \url{https://libntl.org/}.

\bibitem{Crites2020ReputableLC}
Elizabeth~C. Crites, Mary Maller, Sarah Meiklejohn, and Rebekah Mercer.
\newblock Reputable list curation from decentralized voting.
\newblock {\em Proceedings on Privacy Enhancing Technologies}, 2020:297 -- 320, 2020.

\bibitem{sigma_protocol}
Ivan Damg{\aa}rd.
\newblock On $\sigma$-protocols.
\newblock {\em Lecture Notes, University of Aarhus, Department for Computer Science}, page~84, 2002.

\bibitem{device_fingerprint}
Device fingerprinting.
\newblock \url{https://en.wikipedia.org/wiki/Device_fingerprint}.

\bibitem{fiat1986prove}
Amos Fiat and Adi Shamir.
\newblock How to prove yourself: Practical solutions to identification and signature problems.
\newblock In {\em Conference on the theory and application of cryptographic techniques}, pages 186--194. Springer, 1986.

\bibitem{fz_urs}
Eiichiro Fujisaki and Koutarou Suzuki.
\newblock Traceable ring signature.
\newblock In Tatsuaki Okamoto and Xiaoyun Wang, editors, {\em Public Key Cryptography -- PKC 2007}, pages 181--200, Berlin, Heidelberg, 2007. Springer Berlin Heidelberg.

\bibitem{algorand}
Yossi Gilad, Rotem Hemo, Silvio Micali, Georgios Vlachos, and Nickolai Zeldovich.
\newblock Algorand: Scaling byzantine agreements for cryptocurrencies.
\newblock In {\em Proceedings of the 26th Symposium on Operating Systems Principles}, SOSP '17, page 51–68, New York, NY, USA, 2017. Association for Computing Machinery.

\bibitem{Groth2004EfficientMP}
Jens Groth.
\newblock Efficient maximal privacy in boardroom voting and anonymous broadcast.
\newblock In {\em Financial Cryptography}, 2004.

\bibitem{one_many}
Jens Groth and Markulf Kohlweiss.
\newblock One-out-of-many proofs: Or how to leak a secret and spend a coin.
\newblock In Elisabeth Oswald and Marc Fischlin, editors, {\em Advances in Cryptology - EUROCRYPT 2015}, pages 253--280, Berlin, Heidelberg, 2015. Springer Berlin Heidelberg.

\bibitem{ristretto}
Mike Hamburg, Henry de~Valence, Isis Lovecruft, and Tony Arcieri.
\newblock The ristretto group.
\newblock \url{https://ristretto.group/}.

\bibitem{jni}
Java native interface.
\newblock \url{https://en.wikipedia.org/wiki/Java_Native_Interface}.

\bibitem{blind-sigs}
John Kevine and Hassib Ashouri.
\newblock Blind-signatures javascript library on npm.
\newblock \url{https://www.npmjs.com/package/blind-signatures}.

\bibitem{kiayias_et_al:OASIcs:2020:11967}
Aggelos Kiayias, Benjamin Livshits, Andr{\'e}s~Monteoliva Mosteiro, and Orfeas Stefanos~Thyfronitis Litos.
\newblock {A Puff of Steem: Security Analysis of Decentralized Content Curation}.
\newblock In Vincent Danos, Maurice Herlihy, Maria Potop-Butucaru, Julien Prat, and Sara Tucci-Piergiovanni, editors, {\em International Conference on Blockchain Economics, Security and Protocols (Tokenomics 2019)}, volume~71 of {\em OpenAccess Series in Informatics (OASIcs)}, pages 3:1--3:21, Dagstuhl, Germany, 2020. Schloss Dagstuhl--Leibniz-Zentrum fuer Informatik.

\bibitem{cryptoeprint:2022/760}
Aggelos Kiayias, Vanessa Teague, and Orfeas Stefanos~Thyfronitis Litos.
\newblock Privacy preserving opinion aggregation.
\newblock Cryptology ePrint Archive, Paper 2022/760, 2022.
\newblock \url{https://eprint.iacr.org/2022/760}.

\bibitem{kulyk14}
Oksana Kulyk, Stephan Neumann, Melanie Volkamer, Christian Feier, and Thorben Koster.
\newblock Electronic voting with fully distributed trust and maximized flexibility regarding ballot design.
\newblock In {\em 2014 6th International Conference on Electronic Voting: Verifying the Vote (EVOTE)}, pages 1--10, 2014.

\bibitem{moderation-on-reddit-splotlight}
Hanlin Li, Brent Hecht, and Stevie Chancellor.
\newblock All that’s happening behind the scenes: Putting the spotlight on volunteer moderator labor in reddit.
\newblock {\em Proceedings of the International AAAI Conference on Web and Social Media}, 16(1):584--595, May 2022.

\bibitem{linkable_sign}
Joseph~K. Liu, Victor~K. Wei, and Duncan~S. Wong.
\newblock Linkable spontaneous anonymous group signature for ad hoc groups.
\newblock In Huaxiong Wang, Josef Pieprzyk, and Vijay Varadharajan, editors, {\em Information Security and Privacy}, pages 325--335, Berlin, Heidelberg, 2004. Springer Berlin Heidelberg.

\bibitem{Liu2017AnEP}
Yi~Liu and Qi~Wang.
\newblock An e-voting protocol based on blockchain.
\newblock {\em IACR Cryptol. ePrint Arch.}, 2017:1043, 2017.

\bibitem{stackoverflow-motivation}
Yao Lu, Xinjun Mao, Minghui Zhou, Yang Zhang, Zude Li, Tao Wang, Gang Yin, and Huaimin Wang.
\newblock Motivation under gamification: An empirical study of developers’ motivations and contributions in stack overflow.
\newblock {\em IEEE Transactions on Software Engineering}, 48(12):4947--4963, 2022.

\bibitem{May2014FilterF}
Avner May, Augustin Chaintreau, Nitish Korula, and Silvio Lattanzi.
\newblock Filter \& follow: how social media foster content curation.
\newblock In {\em SIGMETRICS '14}, 2014.

\bibitem{anatomy-of-reddit}
Alexey~N Medvedev, Renaud Lambiotte, and Jean-Charles Delvenne.
\newblock The anatomy of reddit: An overview of academic research.
\newblock {\em Dynamics On and Of Complex Networks III: Machine Learning and Statistical Physics Approaches 10}, pages 183--204, 2019.

\bibitem{vrf_1999}
Silvio Micali, Michael Rabin, and Salil Vadhan.
\newblock Verifiable random functions.
\newblock In {\em 40th annual symposium on foundations of computer science (cat. No. 99CB37039)}, pages 120--130. IEEE, 1999.

\bibitem{onion_routing}
Onion routing - {Wikipedia}.
\newblock \url{https://en.wikipedia.org/wiki/Onion_routing}.

\bibitem{reddit-systematic-overeview}
Nicholas Proferes, Naiyan Jones, Sarah Gilbert, Casey Fiesler, and Michael Zimmer.
\newblock Studying reddit: A systematic overview of disciplines, approaches, methods, and ethics.
\newblock {\em Social Media + Society}, 7(2):20563051211019004, 2021.

\bibitem{ring_sign}
Ronald~L. Rivest, Adi Shamir, and Yael Tauman.
\newblock How to leak a secret.
\newblock In Colin Boyd, editor, {\em Advances in Cryptology --- ASIACRYPT 2001}, pages 552--565, Berlin, Heidelberg, 2001. Springer Berlin Heidelberg.

\bibitem{avalanche}
Team Rocket, Maofan Yin, Kevin Sekniqi, Robbert van Renesse, and Emin~G{\"u}n Sirer.
\newblock Scalable and probabilistic leaderless bft consensus through metastability.
\newblock {\em arXiv preprint arXiv:1906.08936}, 2019.

\bibitem{blockene_osdi}
Sambhav Satija, Apurv Mehra, Sudheesh Singanamalla, Karan Grover, Muthian Sivathanu, Nishanth Chandran, Divya Gupta, and Satya Lokam.
\newblock Blockene: A high-throughput blockchain over mobile devices.
\newblock In {\em 14th USENIX Symposium on Operating Systems Design and Implementation (OSDI 20)}, pages 567--582, 2020.

\bibitem{libsodium}
Libsodium documentation.
\newblock \url{https://libsodium.gitbook.io/}.

\bibitem{reddit-political-communities}
Ahmed Soliman, Jan Hafer, and Florian Lemmerich.
\newblock A characterization of political communities on reddit.
\newblock In {\em Proceedings of the 30th ACM Conference on Hypertext and Social Media}, HT '19, page 259–263, New York, NY, USA, 2019. Association for Computing Machinery.

\bibitem{tiktok-invasion-ukraine}
Benjamin Steel, Sara Parker, and Derek Ruths.
\newblock The invasion of ukraine viewed through tiktok: A dataset, 2023.

\bibitem{log_urs}
Anh~The Ta, Thanh~Xuan Khuc, Tuong~Ngoc Nguyen, Huy~Quoc Le, Dung~Hoang Duong, Willy Susilo, Kazuhide Fukushima, and Shinsaku Kiyomoto.
\newblock Efficient unique ring signature for blockchain privacy protection.
\newblock In {\em Information Security and Privacy}, pages 391--407. Springer International Publishing, 2021.

\bibitem{TranNSDI09}
Nguyen Tran, Bonan Min, Jinyang Li, and Lakshminarayanan Subramanian.
\newblock Sybil-resilient online content voting.
\newblock In {\em Proceedings of the 6th USENIX Symposium on Networked Systems Design and Implementation}, NSDI'09, page 15–28, USA, 2009. USENIX Association.

\bibitem{dreaddit-2019}
Elsbeth Turcan and Kathleen McKeown.
\newblock Dreaddit: A reddit dataset for stress analysis in social media.
\newblock {\em arXiv preprint arXiv:1911.00133}, 2019.

\bibitem{content-based}
Donghui Wang, Yanchun Liang, Dong Xu, Xiaoyue Feng, and Renchu Guan.
\newblock A content-based recommender system for computer science publications.
\newblock {\em Knowledge-Based Systems}, 157:1--9, 2018.

\bibitem{Yu_sybil}
Haifeng Yu, Chenwei Shi, Michael Kaminsky, Phillip~B. Gibbons, and Feng Xiao.
\newblock Dsybil: Optimal sybil-resistance for recommendation systems.
\newblock In {\em 2009 30th IEEE Symposium on Security and Privacy}, pages 283--298, 2009.

\bibitem{DBLP:conf/ndss/ZhangOB19}
Bingsheng Zhang, Roman Oliynykov, and Hamed Balogun.
\newblock A treasury system for cryptocurrencies: Enabling better collaborative intelligence.
\newblock In {\em 26th Annual Network and Distributed System Security Symposium, {NDSS} 2019, San Diego, California, USA, February 24-27, 2019}. The Internet Society, 2019.

\bibitem{wikipedia-intrinsic}
Xiaoquan Zhang and Feng Zhu.
\newblock Intrinsic motivation of open content contributors: The case of wikipedia.
\newblock In {\em Workshop on information systems and economics}, volume~10. Citeseer, 2006.

\bibitem{russia-ukraine-reddit-dataset}
Yiming Zhu, Ehsan-ul Haq, Lik-Hang Lee, Gareth Tyson, and Pan Hui.
\newblock A reddit dataset for the russo-ukrainian conflict in 2022.
\newblock {\em arXiv preprint arXiv:2206.05107}, 2022.

\end{thebibliography}
